\documentclass[]{article}

\usepackage[numbers]{natbib}

\usepackage{moreverb,url}

\usepackage[colorlinks,bookmarksopen,bookmarksnumbered,citecolor=red,urlcolor=red]{hyperref}
\usepackage{textcomp}                  
\usepackage{mathptmx}                  
\usepackage{times}                     
\usepackage{tabu}                      
\usepackage{booktabs}                  
\usepackage{microtype}                 
\PassOptionsToPackage{warn}{textcomp}  

\usepackage{helvet}  
\usepackage{courier}  
\usepackage{url}  
\usepackage{graphicx}  
\usepackage{array,multirow}
\usepackage{xcolor}
\usepackage[normalem]{ulem} 
\usepackage{amsfonts}
\usepackage{amsmath}
\usepackage{bm}
\usepackage{amssymb}
\usepackage[export]{adjustbox} 
\usepackage{fancyvrb} 

\newcommand\BibTeX{{\rmfamily B\kern-.05em \textsc{i\kern-.025em b}\kern-.08em
T\kern-.1667em\lower.7ex\hbox{E}\kern-.125emX}}

\title{- Preprint - \\ClustML: A Measure of Cluster Pattern Complexity in Scatterplots Learnt from Human-labeled Groupings } 

\author{Mostafa M Abbas (1,2), Ehsan Ullah (2),  Abdelkader Baggag (2),\\
Halima Bensmail (2), Michael Sedlmair (3) and Micha\"el Aupetit (2)\\\\
(1) Geisinger, Danville, PA, USA\\ 
(2) Qatar Computing Research Institute, HBKU, Doha, Qatar\\
(3) VISUS, University of Stuttgart, Germany\\
Contact: maupetit@hbku.edu.qa
}

\date{}

\begin{document}

\maketitle


\section*{HOW TO CITE}

Please refer to the published version:

ClustML: A Measure of Cluster Pattern Complexity in Scatterplots Learnt from Human-labeled Groupings, Information Visualization Journal 23(2)  105-122 (2024). 
DOI: 10.1177/14738716231220536 

\begin{verbatim}
@article{doi:10.1177/14738716231220536,
author = {Mostafa M Hamza and Ehsan Ullah 
          and Abdelkader Baggag and Halima Bensmail 
          and Michael Sedlmair and Michael Aupetit},
title ={ClustML: A measure of cluster pattern complexity 
    in scatterplots learnt from human-labeled groupings},
journal = {Information Visualization},
volume = {23},
number = {2},
pages = {105-122},
year = {2024},
doi = {10.1177/14738716231220536},
URL = {https://doi.org/10.1177/14738716231220536}
}
\end{verbatim}
DATA: https://paperswithcode.com/dataset/clustme-and-clustml-data-s1-and-s2\\
CODE: https://zenodo.org/records/10208144 DOI: 10.5281/zenodo.10208143

\section*{Abstract}
Visual quality measures (VQMs) are designed to support analysts by automatically detecting and quantifying patterns in visualizations. We propose a new VQM for visual grouping patterns in scatterplots, called ClustML, which is trained on previously collected human subject judgments. 
Our model encodes scatterplots in the parametric space of a Gaussian Mixture Model and uses a classifier trained on human judgment data to estimate the perceptual complexity of grouping patterns. The numbers of initial mixture components and final combined groups. It improves on existing VQMs, first, by better estimating human judgments on two-Gaussian cluster patterns and, second, by giving higher accuracy when ranking general cluster patterns in scatterplots. We use it to analyze kinship data for genome-wide association studies, in which experts rely on the visual analysis of large sets of scatterplots. We make the benchmark datasets and the new VQM available for practical use and further improvements. 

\section{Introduction}

Cluster discovery is a typical task in visual data analysis 
~\cite{Clustrophile2_Cavallo2019,VisualClusterAnalysis_PandeyKFBB16}. Clusters can have various shapes, densities, and other characteristics~\cite{Sedlmair2013_ScatterplotAndDR}, and may exist in different data subspaces.
Fully automated clustering techniques are not always satisfying and might not match the end-user expectations~\cite{PerceptionBasedClustering_AupetitSABB19}. Hence, end-users often use data visualization, usually in scatterplots, to find or validate clusters of interest~\cite{Brehmer_BELIV14}. 

To support these users, several pipelines for visual cluster analysis have been proposed in Visual Analytics~\cite{DimRedClustVA_Wenskovitch2018}.
One way to visually discover clusters in multidimensional (HD) spaces is to use multidimensional projection techniques~\cite{MDPsurveyNonatoAupetit2018}, RadViz~\cite{RadViz_Hoffman1999}, or star coordinate plots~\cite{RadVizVsStarCoord_Rubio2016}. Examining the resulting scatterplots allows for detecting grouping patterns that could support the existence of their multidimensional counterpart. 
But these two-dimensional projections generate artifacts~\cite{MDPsurveyNonatoAupetit2018,RadVizVsStarCoord_Rubio2016}, 
and often one view is not enough to reliably discover all the multidimensional cluster structures \cite{subspaceclustering_TatuMFBSSK12,Elhaik2022}. Moreover, clusters may exist only in subspaces of the data. Hence, visual cluster analysis requires generating projections from possibly many (weighted) combinations of the initial features and different tuning of the parameters of projection techniques~\cite{Clustrophile2_Cavallo2019,iPCA_Jeong2009,ClusterSculptor_BRUNEAU2015,SubspaceVoyager_Wang2017,ProjPursuit_FriedmanTukey74,Buja2004TheoryOD}. Eventually, these techniques allow the analyst to spot the most interesting visual cluster patterns for further investigation. 

With increasing data dimensionality, however, this process often becomes tedious and cumbersome due to the large number of projections to explore visually.
Visual Quality Measures~\cite{VQM_BertiniS06} (VQM) can support users 
in such situations by automatically detecting and quantifying visual patterns~\cite{Wilkinson_scagnostics2005,Matute2018_skeletonScagnostics,qualMetric_Bertini11}. Ranking and arranging visualizations by order of interest concerning a specific type of pattern~\cite{subspaceclustering_TatuMFBSSK12,scagexplorer_Wilkinson2014} lets analysts focus their limited-time budget on the most promising views. We are primarily interested in VQMs for cluster and grouping patterns in our work. Several VQMs have been proposed for that purpose~\cite{Tukey1985,Johansson_clustqualmetric2009,ClustMe_eurovis2019}. Of these, the ClustMe method~\cite{ClustMe_eurovis2019} is based on merging and counting components of a Gaussian Mixture Model (GMM)~\cite{fraley200201} of the points in the scatterplot. It was the first GMM-based VQM for quantifying visual cluster patterns in scatterplots. ClustMe has shown the most accurate performance in ranking human perceptual judgment benchmark data among all competitors. Still, its agreement with these human perceptual rankings is in the $[60\%-80\%]$ range, a relatively low scorewhich we expect to improve by replacing the merging component of ClustMe with a new data-driven model, forming ClustML.

\begin{figure}[h!tbp]
\begin{center}
\includegraphics[width=0.5\linewidth]{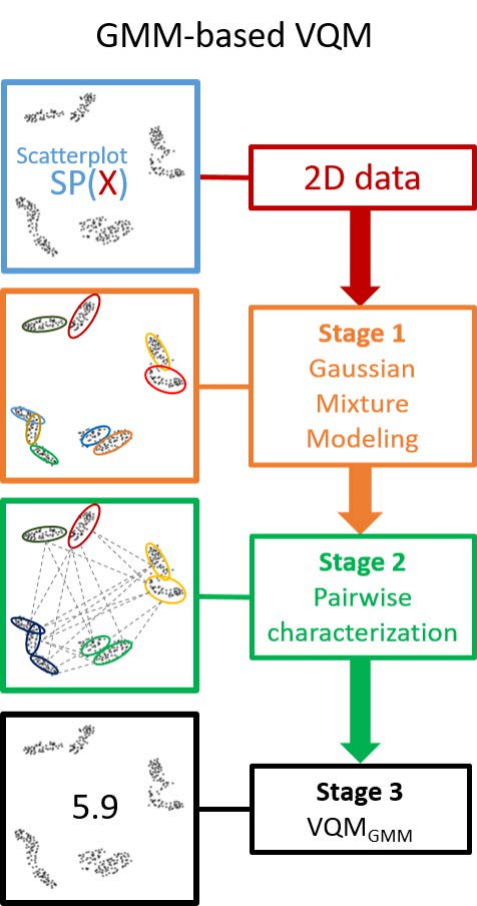}
\caption{\textbf{A visual quality measure (VQM) based on a Gaussian Mixture Model (GMM) for cluster patterns in scatterplots is made of three stages: (1) a data-driven process estimates the parameters of a GMM of the data points density in the scatterplot; (2) the degree of overlapping of each pair of GMM components is computed to provide additional characteristics of interest to quantify cluster patterns; (3) The data points, the GMM parameters, and the pairwise quantities are aggregated to compute the visual quality measure. ClustMe and ClustML are both GMM-based VQMS, differing in the way they quantify pairwise overlap of GMM components (Stage 2).}}
\label{fig:GMM_based_VQM}
\end{center}
\end{figure}

GMM-based VQMs like ClustMe and the proposed ClustML are made of three main stages illustrated in figure \ref{fig:GMM_based_VQM}: 
\begin{itemize}
    \item \textbf{Stage 1: Gaussian Mixture Modeling} of the data points density in the scatterplot. Each Gaussian component of the mixture represents a local subset of the data. The model assumes the data are independently sampled from isolated Gaussian distributions or clusters. A data-driven process estimates the mean, the covariance matrix, and the relative contribution of each component to the global density distribution of the data points.
    \item \textbf{Stage 2: GMM components pairwise characterization} of overlap. When two Gaussian components overlap too much, it is assumed that they likely belong to the same local cluster. Hence, evaluating these pairwise overlaps from the data and the parameters of the GMM provides additional characteristics of interest to quantify cluster patterns. 
    \item \textbf{Stage 3: Visual quality measure computation}. All previous quantities are aggregated to form the final VQM score that quantifies visual cluster patterns in the scatterplot.
\end{itemize}

In ClustMe, the overlapping evaluation (Stage 2) is based on a computational heuristic called \textit{Demp} that decides when two GMM components overlap too much; they are merged or linked together to represent a single cluster instead of two symbolically. The ClustMe VQM score is a linear combination of the number of GMM components and the number of connected components of the graph formed by the \textit{Demp} links, with more weight given to the latter.

In this work, in contrast to ClustMe, we set out to develop ClustML, a new GMM-based VQM whose overlapping evaluation and merging decision (Stage 2)  is \textit{learned from human judgments of cluster patterns} in scatterplots rather than using the \textit{Demp} heuristic. 

We demonstrate the superiority of ClustML against ClustMe, its main competitor, in terms of agreement with human judgments on two perceptual studies datasets, S1 and S2, previously collected for the development and evaluation of ClustMe ~\cite{ClustMe_eurovis2019}:

\begin{itemize}
    \item \textbf{Dataset S1} is a set of binary judgments from $34$ subjects tasked to decide if they can see \textit{one} or \textit{more-than-one} clusters within each of $1000$ scatterplots data generated by sampling two-component GMMs with various parameters. 
    \item \textbf{Dataset S2} is independent of S1. It is a set of ternary judgments from $31$ subjects tasked to decide for each of $435$ pairs of scatterplots if one or the other shows the most complex cluster pattern or if both are equally complex. 
\end{itemize}
    
In the ClustMe paper, S1 is used to select the best merging decision model among a finite set of $7$ heuristics. In contrast, in this work, S1 is used to \textit{train an automatic classifier to mimic human merging decisions}. In both the ClustMe paper and this work, S2 is used to evaluate the resulting GMM-based VQM for a pairwise ranking task. Using the same datasets, S1 and S2 allows a fair and objective comparison between ClustMe and ClustML. 

We also propose qualitative comparisons between ClustMe and ClustML and a usage scenario in the domain of genome-wide association studies (GWAS). In this domain, interesting cluster patterns can be missed because the analysts explore only the scatterplots spanning the leading principal components of the data \cite{Elhaik2022}. We show that ClustML can help detect cluster patterns hidden in subspaces spanned by low-variance principal components without requiring an exhaustive search among all pairs of components. 

Finally, we discuss the challenges in developing hybrid computational-perceptual VQMs for cluster patterns and argue for creating perceptual-study-based benchmark datasets for evaluating and designing new VQMs. 

R codes and datasets S1 and S2 are publicly available \cite{ZENODO_DATA}.


\section{Related work}

We review related work on visual quality measures (VQMs) designed to detect and quantify cluster patterns, VQMs built from data rather than heuristics, and merging decision techniques used in Gaussian Mixture Models specific to our GMM-based VQM approach.

\subsection{Visual quality measure for clustering}
Visual cluster patterns have been taxonomized~\cite{taxonomySepme_SedlmairTMT12} and empirically studied~\cite{VisualClusterAnalysis_PandeyKFBB16}. These works show various characteristics, demonstrating how challenging it is to develop VQMs for such loosely defined pattern types. Several approaches have been proposed to design VQMs for grouping patterns, each focusing on some specific definition. The Clumpiness measure~\cite{Tukey1985} detects clumps in a scatterplot. It is part of the Scagnostics scatterplot descriptors~\cite{Wilkinson_scagnostics2005}. Other VQM approaches are based on CLIQUE clustering~\cite{Johansson_clustqualmetric2009}. 
Existing VQMs are mostly heuristics loosely related to human perceptual data. For instance, Pandey \textit{et al.}~\cite{VisualClusterAnalysis_PandeyKFBB16} showed that Scagnostics are not well-related to their participants' judgments (they were never explicitly designed for that, though). 

In contrast, ClustML is a data-driven VQM directly optimized to mimic human judgments.

\subsection{Data-driven VQMs}

Beyond heuristics-based approaches, data-driven approaches like ScatterNet~\cite{ScatterNet_Ma2018} or perception-based VQMs~\cite{Albuquerque11_PerceptionBasedVQM} get trained on human judgment data.
Recent work on data-driven approaches has shown that fine-tuning a VQM on a specific pattern to mimic \emph{perceptual judgments} outperforms heuristic techniques, for instance, in the case of class separation measures for class color-coded scatterplots~\cite{Sepme1_SedlmairA15,Sepme2_AupetitS16}. These data-driven VQMs led to new applications in supervised dimensionality reduction of labeled data ~\cite{SupDRSepMe_Wang2018} and color optimization for  scatterplots~\cite{OptimColorSepMe_Wang2019}.

Regarding cluster patterns (i.e., no color for class labels in the scatterplot), the X-means~\cite{Pelleg00Xmeans}, DBSCAN~\cite{DBSCAN_1996}, and  CLIQUE~\cite{agrawal1998automaticClustering} clustering techniques, and the Clumpiness~\cite{Wilkinson_scagnostics2005} VQM have been compared to the  ClustMe data-driven VQM~\cite{ClustMe_eurovis2019} on two human judgment benchmark datasets. ClustMe outperformed all others in terms of Vanbelle kappa~\cite{VanbelleKappa_2009} agreement index.    
 
Among all these approaches, only ScatterNet~\cite{ScatterNet_Ma2018} relies on a data-driven parametric model (auto-encoder) of
human judgments rather than a predefined heuristic. Parameters of the model are optimized to predict pairwise similarity judgments between monochrome scatterplots. However, no such model exists for quantifying grouping patterns.

\subsection{GMM-based VQM approach}

Closest to our work is ClustMe~\cite{ClustMe_eurovis2019}, a VQM for grouping patterns based on Gaussian Mixture Models (GMMs).
ClustMe builds a GMM whose components are merged to detect more complex, \emph{non-Gaussian}, grouping patterns (See~\cite{Bensmail1997,fraley200201} for an overview). Human-subject data S1 has been used to evaluate and select the best merging criterion (\textit{Demp})  among seven heuristics~\cite{mergingGMM_Hennig10}, resulting in $60\%$ to $80\%$ agreement between ClustMe and human perceptual judgments. However, these heuristics are designed by data analysts grounded on mathematical principles rather than directly from perceptual judgments. A more recent work \cite{Jeon2023CLAMS} uses an approach similar to ClustMe but considers cluster ambiguity measured with Shannon entropy of human judgments S1 instead of cluster separation. It also uses feature engineering to generate various aggregate heuristics of the GMM parameters and analyze factors at play in visual perception of cluster ambiguity.  Due to the success of machine learning approaches in many domains, we hypothesized that applying such an approach instead of heuristics or feature engineering could benefit GMM-based VQM. 

ClustML uses the same GMM-based VQM architecture as ClustMe (\autoref{fig:ClustML_pipeline}(a)); however, human perceptual judgment in dataset S1 are directly used to model the merging decision function by training an automatic classifier (\autoref{fig:ClustML_pipeline}(b)). As a result, the merging model in ClustML reaches more than $96\%$ agreement (\emph{almost perfect agreement}) with human-judgment evaluation data.  
This study presents the detailed architecture and training process of the merging function that makes ClustML outperform ClustMe on a second human-judgment benchmark dataset~\cite{ClustMe_eurovis2019} S2 designed to evaluate VQMs by ranking scatterplots based on their grouping patterns.


\section{ClustML: principle and design}
\label{sec_clusteringpipeline}
\begin{figure}[h!tbp]
\begin{center}
\includegraphics[width=1\linewidth]{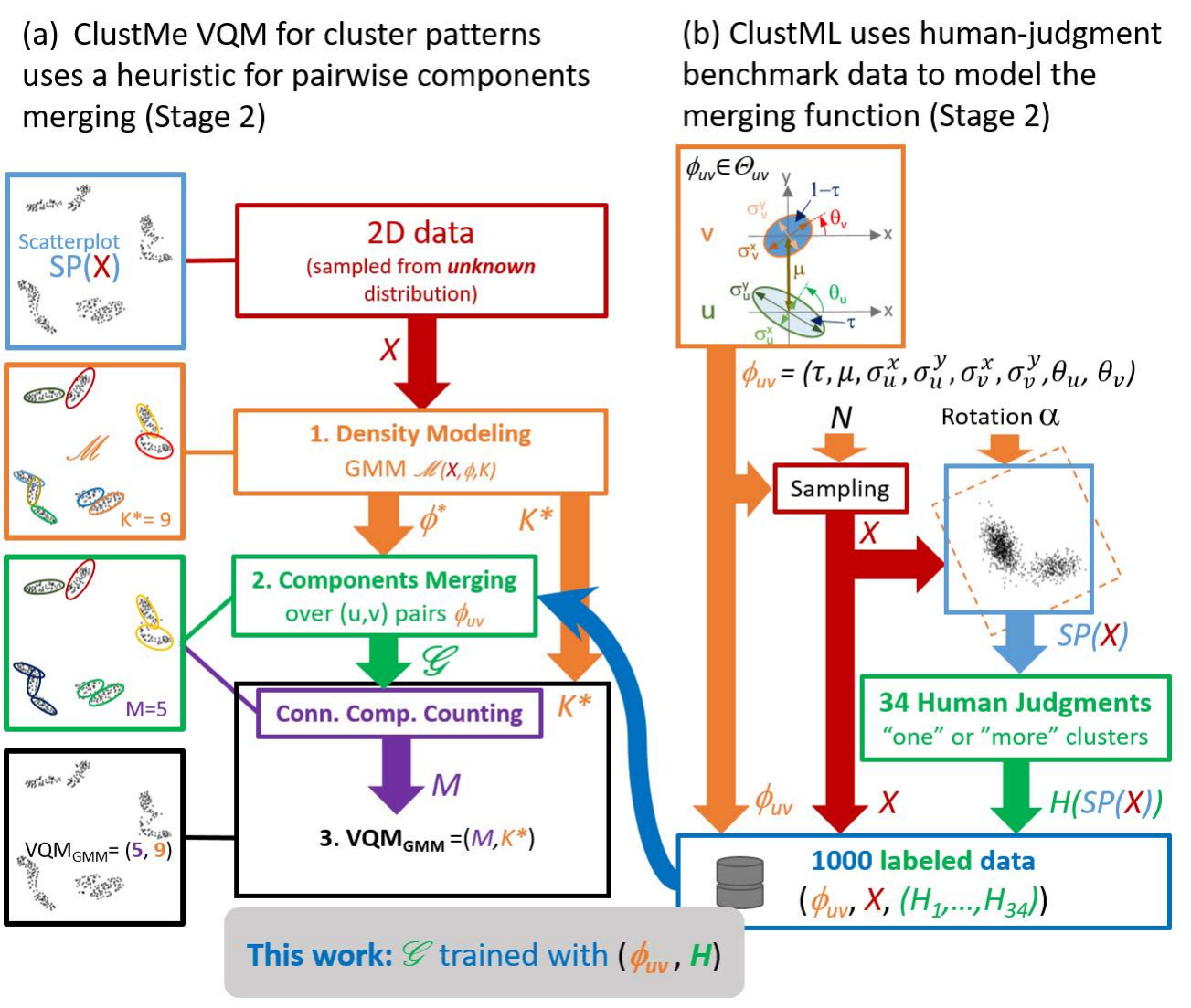}
\caption{\textbf{ClustMe and ClustML are GMM-based VQMs for cluster patterns.} (a) The VQM pipeline of  ClustMe uses a heuristic (\textit{Demp}) as a merging decision function for each pair of GMM components. (b) ClustML follows the same pipeline as ClustMe but uses an automatic classifier as a merging decision function (green) trained on $1000$ monochrome scatterplots from a previous study~\cite{ClustMe_eurovis2019}. These scatterplots were generated in study S1 from varying the parameters $\phi_{uv}$ of a GMM with 2 components and labeled by $34$ subjects $(H_1,...,H_{34})$ seeing \textit{one} ($H_n=0$) or \textit{more-than-one} ($H_n=1$) clusters.}
\label{fig:ClustML_pipeline}
\end{center}
\end{figure}

We give a more technical view of the ClustMe pipeline (\autoref{fig:ClustML_pipeline}(a)), then we present the main principle of ClustML's merging function and its pre-processing and training protocols on data S1 (\autoref{fig:ClustML_pipeline}(b)).

\subsection{ClustMe VQM for grouping patterns}

In the following, we consider a set of $N$ data points ${X}=\{{x}_1,\dots,{x}_N\}\in(\mathbb{R}^2)^N$ in a $2$-dimensional real space, represented graphically as a scatterplot $SP(X)$. 

The previously proposed ClustMe~\cite{ClustMe_eurovis2019} follows the three GMM-based VQM stages illustrated in \autoref{fig:ClustML_pipeline}(a): 

\begin{itemize}
\item \textbf{Stage 1: Gaussian Mixture Modeling.} 
The probability density of the data points is modeled with a Gaussian Mixture Model $\mathcal{M}(X,{\phi},K)$~\cite{fraley200201,Bensmail1997} with $K$ bivariate Gaussian distribution components $g$. The probability density at any point ${x}\in {\mathbb{R}^2}$ is estimated given  model parameter ${\phi}=(\pi_1,\dots,\pi_K,{\mu}_1,\dots,{\mu}_K,{\Sigma}_1,\dots,{\Sigma}_K)$ by: 

\begin{equation}\label{eqDensModel}
p({x}|{\phi},K)=\sum_{k=1}^K \pi_k g({x},{\mu}_k,{\Sigma}_k)
\end{equation}
with 
$g({x},{\mu},{\Sigma}) \!=\! \det(2\pi{\Sigma})^{-\frac{1}{2}}e^{-\frac{1}{2}({x}-{\mu})^\top{\Sigma}^{-1}({x}-{\mu})}$ and  $\sum_k\pi_k = 1$.

The parameter vector $\phi_K$ controls the location $\mu_k$, shape $\Sigma_k$ and weight $\pi_k$ of each component of $\mathcal{M}$. The Bayesian Information Criterion defined by Schwarz~\cite{schwarz78_BIC} is maximized to determine the best model $\mathcal{M}^*$, with the number of components $K^*$ and parameter ${\phi}^*$. 

\item \textbf{Stage 2: GMM components pairwise characterization.} 

Each pair of components $(u,v)\in \{1,\dots,K^*\}^2, u\neq v$ of $\mathcal{M}^*$ is independently screened by the \textit{Demp}  merging heuristic $\mathcal{G}_{\textit{Demp}}$ to decide if it forms a single cluster locally. 
$\mathcal{G}_{\textit{Demp}}({X},{\phi}^*,u,v)\in{\{0,1\}}$ takes the binary decision to merge (1) or not (0) the two components $u$ and $v$ based on the optimal parameter $\phi^*$ and data $X$.

\item \textbf{Stage 3: Score computation}
The adjacency matrix $\left(\mathcal{G}\right)_{u,v}$ forms a graph whose vertices are the $K^*$ components of $\mathcal{M}^*$ and $M\;(\leq K^*)$ its number of connected components. 
Finally, the pair $VQM_{\textit{ClustMe}}(X) = (M,K^*)$ quantifies the complexity of the visual cluster pattern in $SP(X)$: scatterplots $SP(X_i)$ are ranked first by $M_i$ then by $K_i^*$ for equal $M_i$. 
In other words, ClustMe tells that a scatterplot $SP(X_h)$ displays a more complex cluster pattern than $SP(X_l)$ if $$M_l<M_h \;\textrm{or}\; (M_l = M_h\;\textrm{and}\;K^*_l<K^*_h )$$

$\displaystyle VQM_{\textit{ClustMe}}(X_i) = M_i + \frac{K_i^*}{1+K^*_{max}}$ can be used instead, with $K^*_{max}$ the maximum number of GMM components obtained across all $SP(X_i)$ to be compared.

\end{itemize}

In contrast to ClustMe~\cite{ClustMe_eurovis2019}, the main idea of  ClustML is to use an automatic \emph{binary classifier} \textit{trained on human judgment data} to realize the \textit{merging function} $\mathcal{G}_{\textit{ClustML}}$ instead of  $\mathcal{G}_{\textit{Demp}}$ in Stage $2$. All other processes in the above stages are identical for ClustMe and ClustML. However, for the same data $X$ and GMM $\mathcal{M}^*$ (Stage 1), the different $\mathcal{G}_{\textit{ClustML}}$ merging function (Stage 2) can lead to a different value of $M$ and finally, a different $VQM_{ClustML}$ score (Stage 3). We detail the design of $\mathcal{G}_{\textit{ClustML}}$ and its training protocol in the next sections.

\subsection{ClustML merging from human judgment data}

\begin{figure}[h!tbp]
\begin{center}
\begin{tabular}{ccl}

\multirow{4}{*}{
\includegraphics[width=0.75\textwidth]{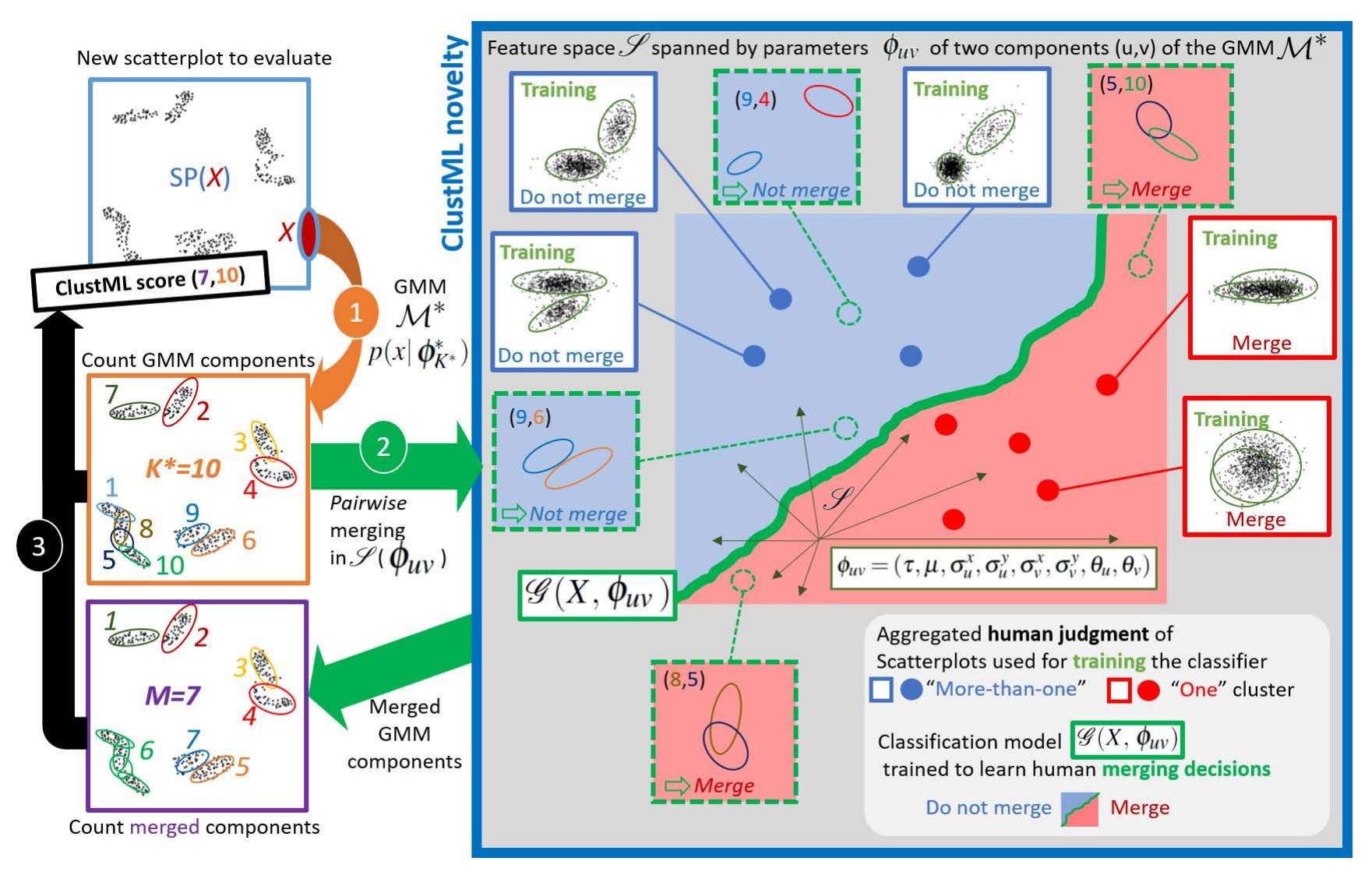}}
&&\\  
&\includegraphics[width=0.1\textwidth,cfbox=blue 1pt 0pt]{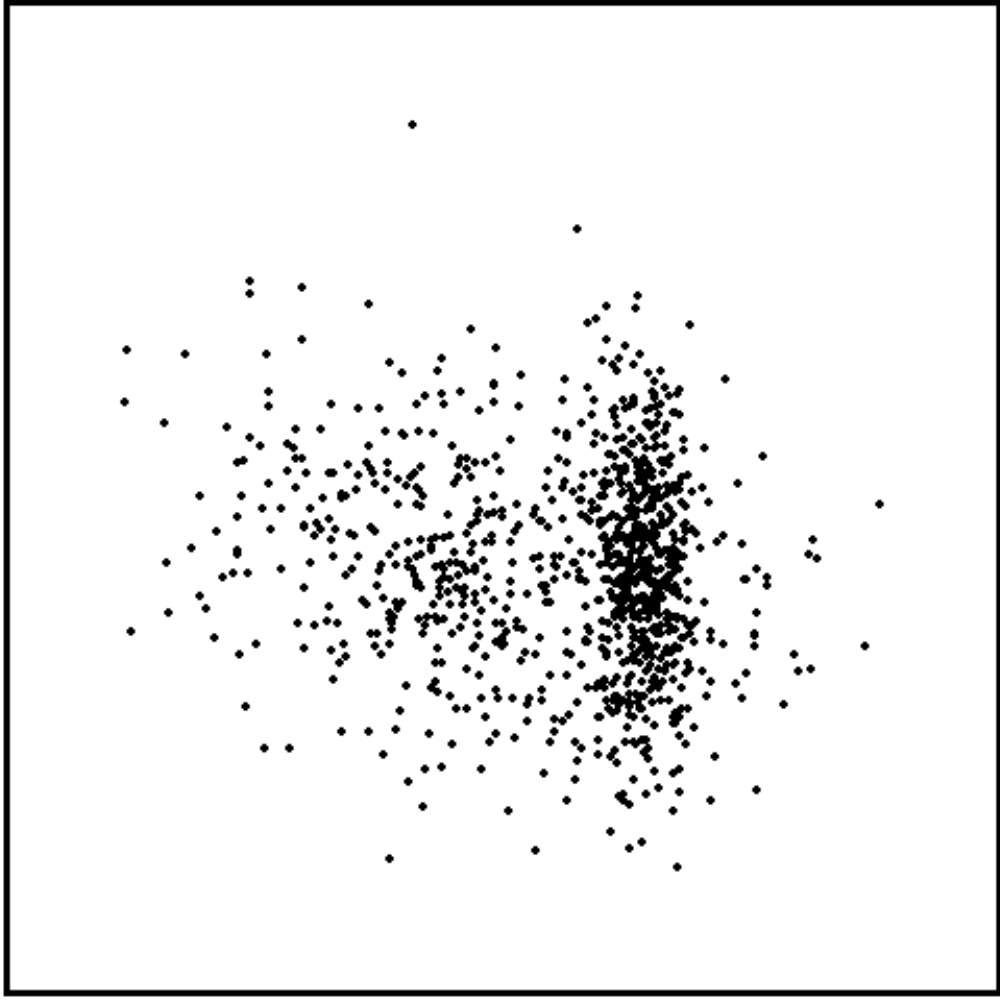}
&\small$64.7\%$
\\
&\includegraphics[width=0.1\textwidth,cfbox=blue 1pt 0pt]{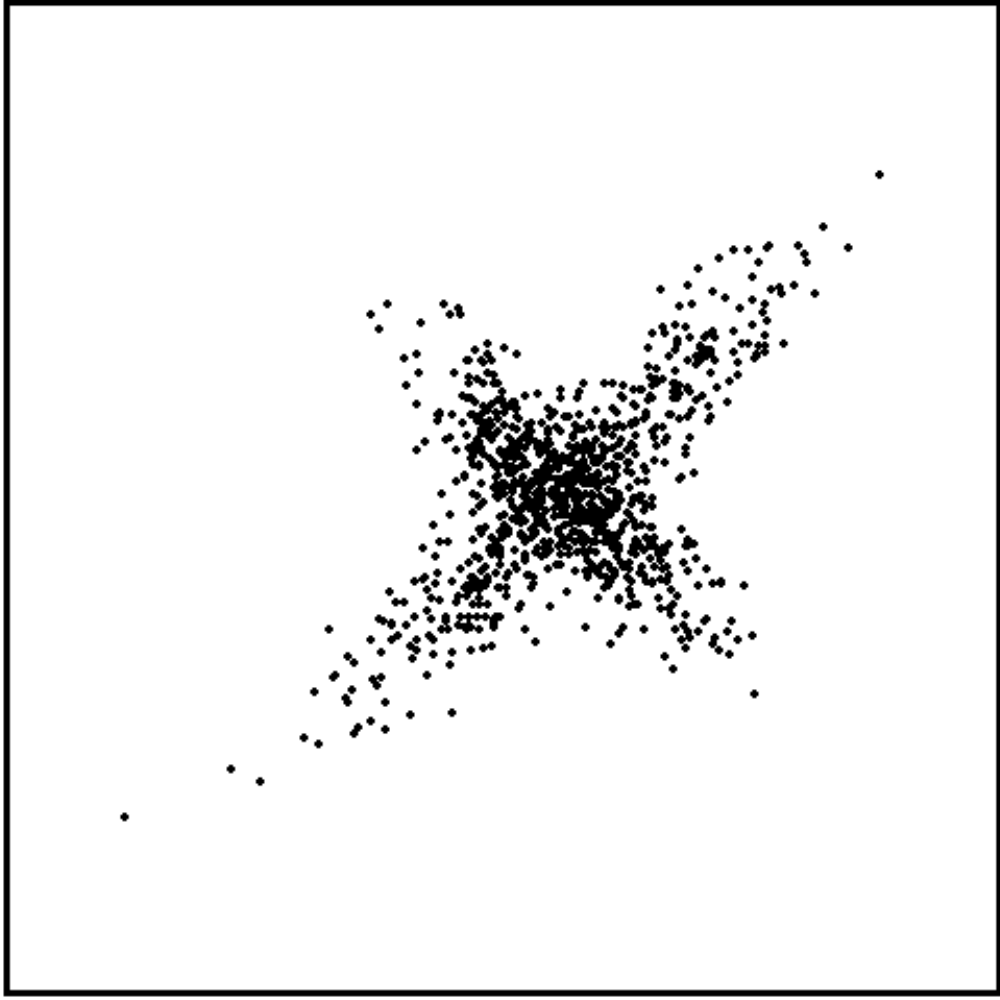}
&\small$58.8\%$
\\
&\includegraphics[width=0.1\textwidth,cfbox=red 1pt 0pt]{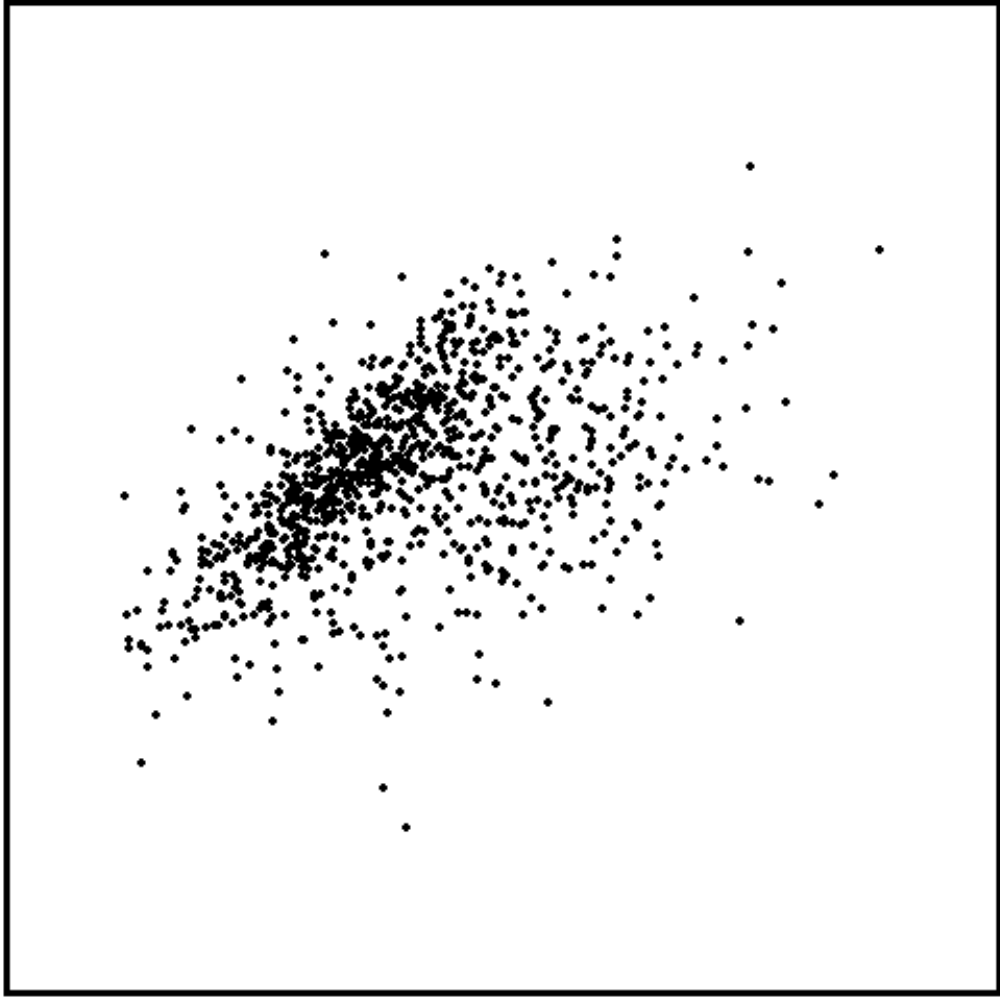}
&\small$17.6\%$
\\
&\includegraphics[width=0.1\textwidth,cfbox=red 1pt 0pt]{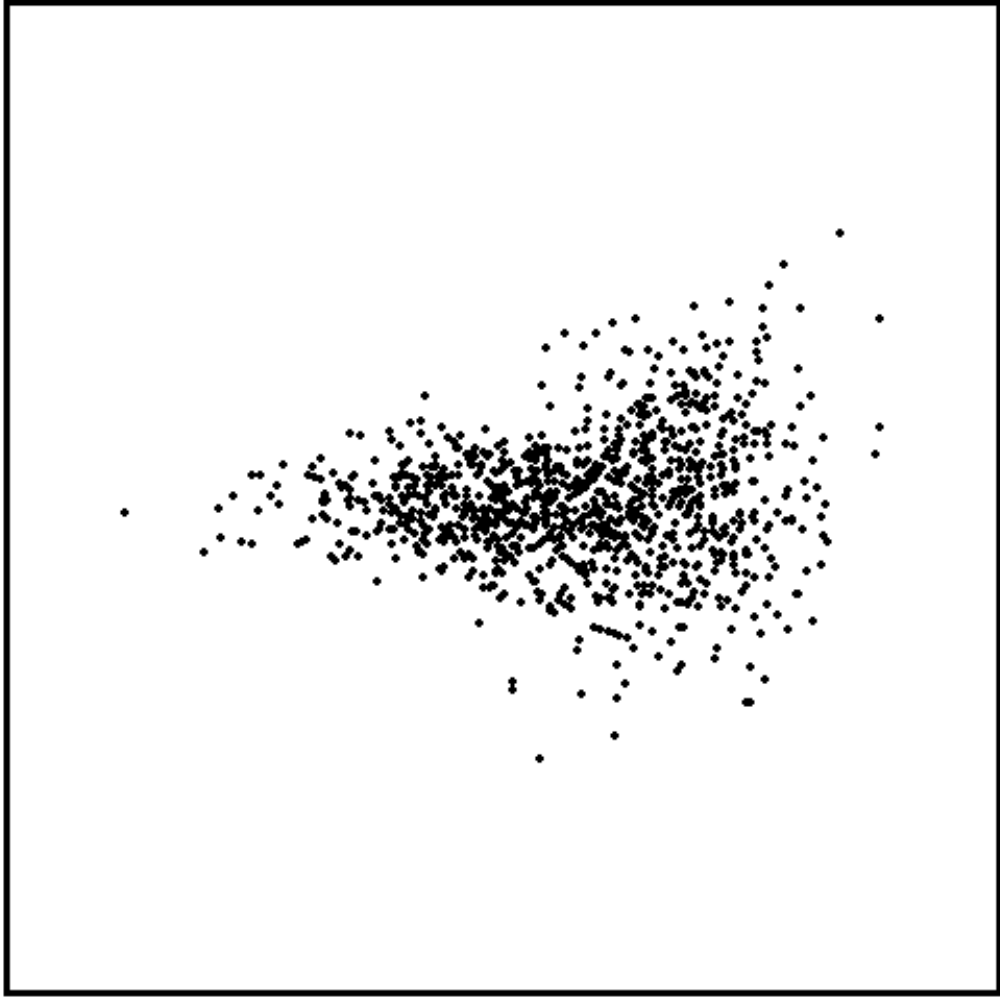}
&\small$2.9\%$
\\
\end{tabular}

\caption{\textbf{ClustML measures the amount of grouping in scatterplots based on a classifier trained on human judgments:} A bivariate Gaussian Mixture Model (Stage 1) models the distribution of the points in the scatterplot to evaluate. Each possible pair of its $K^*$ Gaussian components is assessed for merging (Stage 2). For that purpose and as the main novelty of that work, a binary classifier $\mathcal{G}$ has been trained in the parameter space $\Phi_{uv}$ of component pairs $(u,v)$ (red and blue dots on the right; actually, this space has $8$ dimensions). Scatterplots (Solid red and blue frames) generated by $1000$ pairs have been labeled in a previous experiment~\cite{ClustMe_eurovis2019} by $34$ subjects tasked to decide whether each scatterplot shows \textit{one} (\textcolor{red}{Red}) or \textit{more-than-one} (\textcolor{blue}{Blue}) clusters. Five such ``Training'' scatterplots with plain-line blue or red frames are displayed, and four others in the right column with the percentage of subjects seeing \textit{more-than-one} cluster. After training, the classifier $\mathcal{G}_{ClustML}$ automatically predicts the merging decision (Green solid line separating blue and red areas) that humans would take for yet unseen $2$-Gaussian scatterplots (Dashed green frames). This GMM component pairwise merging decision generates a set of $M$ connected components (purple frame). Finally, the ClustML VQM (Stage 3) of the evaluated scatterplot is given by the pair $(M,K^*)$; the higher the score, the more complex the grouping pattern.}
\label{fig:ClustML_overview}
\end{center}
\end{figure}
 
 The scatterplot stimuli used in study S1 of  ClustMe~\cite{ClustMe_eurovis2019} were generated from a bivariate GMM made of two ($K=2$) Gaussian components $u$ and $v$ (\autoref{fig:ClustML_pipeline}b), varying parameters $\phi_{uv}$.  
 The parameter space $\mathcal{S}$ spanned by the vectors $\phi_{uv}$ 
contains all possible mixtures of two bivariate Gaussian distributions. A point $\phi_{uv}^{[i]}$ in that space determines a unique mixture distribution $\mathcal{M}_{\phi_{uv}^{[i]}}$ from which one can randomly sample $N$ points $X^{[i]}$ to generate a 2D scatterplot $SP(X^{[i]})$ with a unique cluster pattern up to sampling variation. 
As illustrated in Figure \ref{fig:ClustML_overview}, in some regions of this multidimensional parameter space $\mathcal{S}$, the generated scatterplots will show two clearly  separated Gaussian clusters (top left blue area), while in other regions the scatterplots will show a single blob of two strongly overlapping Gaussian distributions (bottom right red area). How can we decide about merging two components $u$ and $v$ depending on the position in that space, \textit{i.e.} depending on the values of the parameter $\phi_{uv}$?  

Based on S1 data, we can assign label $1$ or $0$ to a vector $\phi_{uv}^{[i]}$ for which most participants judged the scatterplot $SP(X^{[i]})$ was showing  \emph{one} or \emph{more-than-one} clusters respectively. Then $1$ codes for the \emph{merge} decision, while $0$ codes for the \emph{do-not-merge} decision. ClustML uses such data $\phi_{uv}^{[i]}$ to train a \textit{binary classifier} $\mathcal{G}_{\textit{ClustML}}$ in the space $\mathcal{S}$ to model this human judgment. Finally, the classifier computes a merging decision $\mathcal{G}_{\textit{ClustML}}(X,\phi,u,v)$ for any possible pair $(u,v)$ of components in $\mathcal{M}$ projected in $\mathcal{S}$. 

 We follow the below protocol to train this classifier:

\begin{enumerate}
\item \textbf{Summarize human judgments} from the study S1 to form the labeled dataset $\mathcal{X}_{uv}$;
\item \textbf{Align space} $\mathcal{S}$ of the dataset $\mathcal{X}_{uv}$ with the parameter space of the density model obtained at Stage 1;
\item \textbf{Augment the dataset} $\mathcal{X}_{uv}$ to ensure better generalization of the classifier $\mathcal{G}_{ClustML}$; 
\item \textbf{Train the classifier} $\mathcal{G}_{ClustML}$ with the augmented data to get the optimal merging decision at Stage 2.  

\end{enumerate}

Now, we justify and detail each step of this protocol.

\subsection{Summarizing human judgments} 
\label{sec:summary_human_judgments}

Study S1 gives several human judgments for each scatterplot. Still, we need a single judgment (class label) per scatterplot to train a binary classifier, so we summarize these judgments using a majority vote in the following way. 

We form the labeled dataset $\mathcal{X}_{uv}=\{(input,label)\}_i=\{(\phi_{uv}^{[i]},H^{[i]})\}_{i}$ by pairing the summary $H^{[i]}$ of $34$ perceptual judgments $(H_1^{[i]},\dots,H_{34}^{[i]})$ of cluster patterns in scatterplots stimuli $SP(X^{[i]})$ collected from study S1 together with the parameters $\phi_{uv}^{[i]}$ of the 2-dimension 2-component GMM from which were sampled the points $X^{[i]}$.
We summarize the $34$ human judgments into a binary class $H^{[i]}\in\{0,1\}$ by applying a majority vote. Label $H^{[i]}=0$ (\emph{do not merge}) is assigned to input $\phi_{uv}^{[i]}$ if most of the judgments on $SP(X^{[i]})$ are \emph{more-than-one} cluster. Label $H^{[i]}=1$ (\emph{merge}) is assigned otherwise.
A training data $\chi_i$ is a pair $(\phi_{uv}^{[i]},H^{[i]})\in\mathcal{X}_{uv}$. We note $\Phi_{uv} = \{\phi_{uv}^{[i]}\}_i$ the unlabeled part of these data.

\subsection{Parameter space alignment} 

 The space $\mathcal{S}$ spanned by vectors $\Phi_{uv}$ of the GMM used to generate scatterplots in S1 (\ref{fig:ClustML_overview}) does not match with the space spanned by the parameters $\phi$ of the GMM (Stage 1) of  $X$. We need to transform $\phi$ into $\phi_{uv}$ to get the labeled data $\mathcal{X}_{uv}$.

\begin{figure}[h!tbp]
\begin{center}
\includegraphics[width=0.4\linewidth]{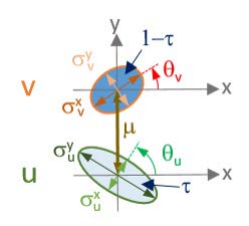}
\caption{Parameters ${\phi}_{uv}=(\tau,\mu,\sigma_u^x,\sigma_u^y,\sigma_v^x,\sigma_v^y,\theta_u,\theta_v)$ of a pair of Gaussian components $(u,v)$ of $\mathcal{M}^*$ control the direction ($\theta$), the probability ($\tau$), the extent ($\sigma$), and the distance ($\mu$) of the two component distributions, hence the (perceptual) overlap of their sampled data. These parameter vectors span the feature space $\mathcal{S}$ (Figure \ref{fig:ClustML_overview}) input of the classifier $\mathcal{G}_{ClustML}$ taking decision of merging $u$ and $v$. }
\label{fig:ClustML_parameterUV}
\end{center}
\end{figure}

Consider a single pair $(u,v)$ of components of $\mathcal{M}^*$. 
The space spanned by the parameters related to $u$ and $v$ only, ${\phi}_{uv}=(\pi_u,\pi_v,\mu_u,\mu_v,\Sigma_u,\Sigma_v)\subseteq \phi$ has a \textit{fixed dimension} ($|{\phi}_{uv}|=14$) independent of $K^*$, which makes it suitable for standard vector-based machine learning.
${\phi}_{uv}$ can be further reduced to a set of $8$ independent parameters (Figure \ref{fig:ClustML_parameterUV}) due to cross-dependencies:
\begin{equation}
\label{eq:case_param_uv}
{\phi}_{uv}=(\tau,\mu,\sigma_u^x,\sigma_u^y,\sigma_v^x,\sigma_v^y,\theta_u,\theta_v)\in[0,1]\times (\mathbb{R}^+)^5\times[0,\pi/2]^2
\end{equation}

where $\tau=\pi_u/(\pi_u+\pi_v)$, and $\mu=||\mu_v-\mu_u||$. In the sequel, $\mathcal{S}$ is the space spanned by these $8$-dimension vectors ${\phi}_{uv}$.

Following ~\cite{Bensmail1997}, the parameters $\sigma$ and $\theta$ in ${\phi}_{uv}$ come from the Singular Value Decomposition of the covariance $\Sigma_i$ ($i\in\{u,v\}$) into the diagonal "scaling" matrix of eigenvalues $S_i$ and the "rotation" matrix of eigenvectors $R_i$: $\Sigma_i=R_iS_i^2R_i^T$. $S_i$ is a diagonal scaling matrix with independent scales $\sigma_i^x$  and $\sigma_i^y$  along $x$ and $y$ orthogonal axes respectively. This gives an elliptic shape to the mixture components with width and length driven by $x$ and $y$, whenever $\sigma_i^x\neq \sigma_i^y$. $R_i$ is a rotation matrix of angle $\theta_i$ which orients the elliptic shape with respect to the $x$-axis: 
%
\begin{equation}
\label{eq:var_rot_scale_decomp}
   S_i=
  \left( {\begin{array}{cc}
   \sigma_i^x  & 0 \\
   0 & \sigma_i^y 
  \end{array} } \right) \quad \quad   
  R_i=
  \left( {\begin{array}{cc}
   \cos\theta_i & -\sin\theta_i \\
   \sin\theta_i & \cos\theta_i 
  \end{array} } \right)
\end{equation}

The data $X$ of any scatterplot in study S1 were \emph{generated} with a GMM by specifying rotation ($\theta_i\in[0, \pi/2]$)  and scaling ($\sigma_{i}^{x,y}$) parameters to get the covariance matrix $\Sigma_i=R_iS_i^2R_i^T=f(\theta_i,\sigma_i^x,\sigma_i^y)$ (see Equation \ref{eq:var_rot_scale_decomp}, Table \ref{tab:Parameters}, Figure \ref{fig:ClustML_parameterUV}). Let's consider a scatterplot $SP(Y)$ to be scored with ClustML, and $\mathcal{M}^*(Y)$ the best GMM  modeling the density of its points $Y$. The estimated covariance matrix $\hat{\Sigma}_i$ of each component $i$ of $\mathcal{M}^*(Y)$ must be decomposed using SVD into $\hat{S}_i$ and $\hat{R}_i$ (\ref{eq:var_rot_scale_decomp}) from which we get angle $\hat{\theta_i}$ and scaling parameters $\hat{\sigma}_{i}^{x,y}$. Unfortunately, the estimated angle $\hat{\theta}_i$ lies in the range $[-\pi/2,\pi/2]$. In order to align angles $\theta_i$ of training data with estimated angles $\hat{\theta}_i$, we passed each triplet $(\theta_i,\sigma_i^x,\sigma_i^y)$ of all training data $\phi_{uv}$ into the SVD composition-decomposition process: $({\theta'}_i,{\sigma'}_i^x,{\sigma'}_i^y)=SVD(f(\theta_i,\sigma_i^x,\sigma_i^y))$.

Moreover, given the points $Y$ of a new scatterplot, the optimal parameters $\phi^*_{uv}=\{\tau, \mu,\sigma^x_u, \sigma^y_u, \sigma^x_v, \sigma^y_v,\theta_u,\theta_u\}$, obtained from a pair $(u,v)$ of components of the best model $\mathcal{M}^*(Y)$, need to be \emph{scaled}. Indeed, it is likely that the scale of the points $Y$ is orders of magnitude bigger or smaller than the one of the points $X$ in S1's scatterplots. This scaling factor impacts parameters $\mu$ and $\sigma$. 
 We must also correct the angles $\theta_u$ and $\theta_v$ defined relatively to the axis orthogonal to $(\mu_u-\mu_v)$. At the same time, the rotation matrix $R_i$ of the SVD decomposition of inferred $\Sigma_i$ is relative to the vector space of the points $Y$.
Therefore, for any parameter vector $\phi_{uv}^*$ inferred from a \textit{new} scatterplot $SP(Y)$, we first compute the correcting angle $\beta=\angle (\overrightarrow{\mu_u\mu_v},\overrightarrow{y})$ between the two components' centers and the y-axis of points $Y$. We add $\beta$ to all $\theta$ angles. Then we rescale the parameters $\sigma^x_u,  \sigma^y_u, \sigma^x_v, \sigma^y_v$ and $\mu$ by dividing them by the maximum of these values $s=\max(\{\mu,\sigma^x_u, \sigma^y_u, \sigma^x_v, \sigma^y_v\})$. 

Finally, we obtain the input data $\phi^{*align}_{uv}$ to the merging function (classifier) $\mathcal{G}_{ClustML}$:

\begin{equation}
\label{eq:phi_align}
\phi^{*align}_{uv}=align(\phi^*_{uv})=(\tau,\frac{\mu}{s},\frac{\sigma^x_u}{s},\frac{\sigma^y_u}{s},\frac{\sigma_v^x}{s},\frac{\sigma_v^y}{s},\theta_u+\beta,\theta_v+\beta)
\end{equation}

Regarding \emph{training data} $\Phi_{uv}$ from study S1, we first apply the composition-decomposition process $SVD \circ f$ to get $\theta'_i$, then we rescale $\mu$ and $\sigma$. However, correcting the rotation by $\beta$ is useless as the y-axis is, by definition, directed by the  components' centers ($\beta=0$) (See Figure \ref{fig:ClustML_parameterUV}):

\begin{equation}
\begin{split}
\label{eq:case_final_scaled_training_set}
\mathcal{X}^{align}_{uv}=\left\{(\tau,\frac{\mu}{s},\frac{{\sigma'}_u^x}{s},\frac{{\sigma'}_u^y}{s},\frac{{\sigma'}_v^x}{s},\frac{{\sigma'}_v^y}{s},{\theta'}_u,{\theta'}_v,H_i)\right.\\
|\;(\tau,\mu,\sigma^x_u,\sigma^y_u,\sigma_v^x,\sigma_v^y,\theta_u,\theta_v,H_i)\in\mathcal{X}_{uv},\\ ({\theta'}_j,{\sigma'}_j^x,{\sigma'}_j^y)=SVD(f(\theta_j,\sigma_j^x,\sigma_j^y)),\; \forall j\in\{u,v\},\\
\left.s=\max(\{\mu,{\sigma'}^x_u, {\sigma'}^y_u, {\sigma'}^x_v, {\sigma'}^y_v\})\right\}
\end{split}
\end{equation}

The data set $\mathcal{X}^{align}_{uv}$ forms the aligned data to be augmented before training the classifier $\mathcal{G}_{ClustML}$.

\subsection{Data augmentation}

The way we parameterize the pairs of Gaussian components and the way the data S1 were generated lead to a possible lack of data to cover $\mathcal{S}$ sufficiently and get a more generalizable classifier $\mathcal{G}_{ClustML}$.

Data augmentation~\cite{ImgDataAugment_Shorten2019} is a process to enrich the data space with new data in areas where they are lacking to ensure a better prediction by the model, but without requiring additional human labeling. It relies on symmetries to justify that existing labeled data can be replicated in other places of the data space.

\begin{figure}[h!tbp]
\centering
\begin{tabular}{c}
\includegraphics[width=0.95\textwidth]{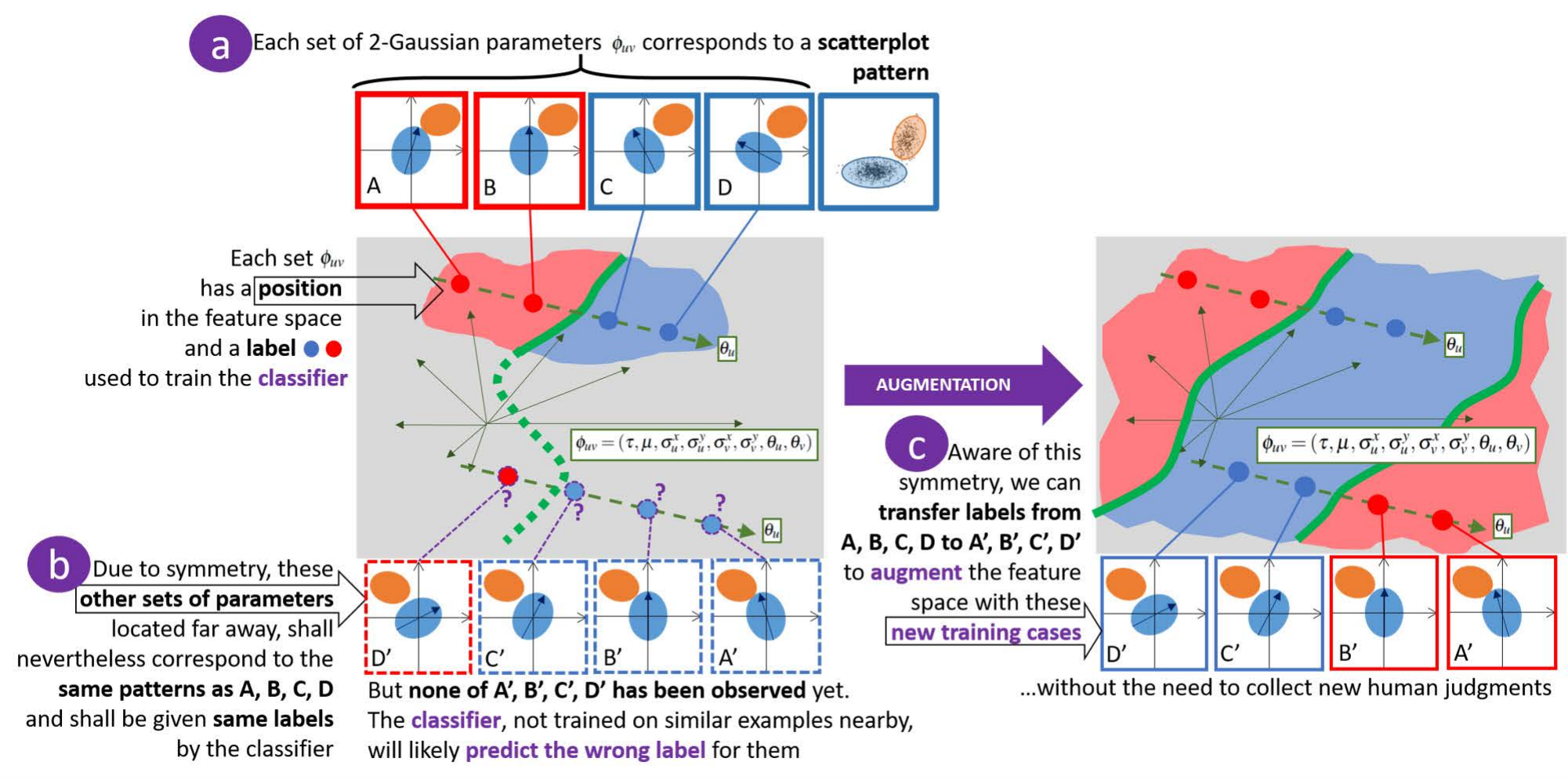}\\
\includegraphics[width=0.95\textwidth]{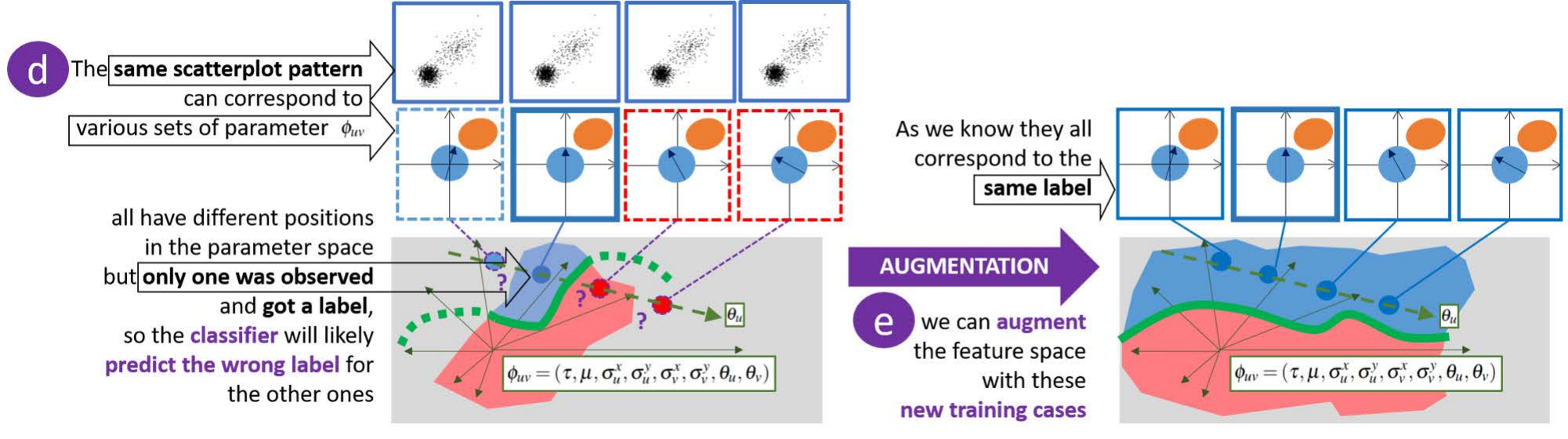}
\end{tabular}
\caption{\textbf{Data augmentation process}: (a) We expect that each set of parameters of a pair of GMM components corresponds to a unique scatterplot up to the sampling variability and vice-versa. But there are symmetries for some settings of these parameters or some scatterplots. (b) Parameters of a pair of components (A, B, C, D) can be different while they represent the exact same cluster pattern in the scatterplot respectively (A', B', C', D') due to symmetry or rotation of the group of points in the scatterplot. (c) Data augmentation involves exploiting these known symmetries to generate additional data (A', B', C', D') with labels corresponding to their symmetrical version (A, B, C, D), enriching the dataset and improving classifier generalizability.
(d) GMMs can model the same scatterplot with different parameters, leading to different locations in the feature space. (e) We generate new data in the feature space leading to the same scatterplot, hence the same label. In all cases (c, e), we need to cover the feature space with labeled examples better to support the training of the classifier; otherwise, the classifier will generalize poorly in these areas (Left side, b, d). The human judgment dataset S1 does not contain such symmetries because it has been designed to avoid showing twice the same scatterplot to human subjects. Therefore, we need to augment these data in the feature space by duplicating labeled scatterplots considering these symmetries (Right side, c, e).}
\label{fig:ClustML_param_simplif}
\end{figure}

We consider the symmetries arising in the parametric representation of a pair of Gaussian components $(u,v)$ in $\mathcal{S}$ (see Figure \ref{fig:ClustML_param_simplif}). Indeed, the parameters used to generate the data sample in study S1 were intended to fall into a restricted part of $\mathcal{S}$ to avoid the same stimuli being shown to the participant while different random parameters are generated. For instance, the scatterplot $SP(\phi_{uv})$ generated by ${\phi_{uv}}=(\tau,\dots,\theta_u,{\theta}_v)$ is identical up to sampling variation, to the one generated by ${\phi_{uv}'}=(\tau,\dots,\theta_u,{\theta}_v+\pi)$ despite ${\phi_{uv}}\neq{\phi_{uv}'}$.

In contrast, we need to cover extensively the parameter space $\mathcal{S}$  with labeled data to get the best possible generalization from the classifier $\mathcal{G}_{ClustML}$, i.e., predicting accurately human judgments for yet unseen scatterplots. Indeed, two scatterplots with perceptually very similar point distributions $X_A$ and $X_B$ will likely get the same human judgment $H_A=H_B$. However, $\phi_A$ can end up very close in $\mathcal{S}$ to a training data $\phi_{T}\in \Phi_{uv}$,  while $\phi_B$ can end up far from it due to the inference process to get $\mathcal{M}^*$. Thus, a classifier trained on $(\phi_T,H_T)$ will be able to predict $H_A\approx H_T$ but will not be good at predicting $H_B$. Therefore, we propose to augment the data $\mathcal{X}_{uv}$ by \emph{replicating} some of the training data $\chi_i=(\phi_i,H_i)$ in different locations $\phi_i'$ of $\mathcal{S}$ to better cover it, getting new data $\chi_i'=(\phi_i',H_i)$ with same label.

For any aligned data $\chi_i=(\phi_i,H_i)\in \mathcal{X}^{align}_{uv}$: 
\begin{equation}
\label{eq:case_0}
{\chi}_i=(\tau,\mu,\sigma_u^x,\sigma_u^y,\sigma_v^x,\sigma_v^y,\theta_u,\theta_v,H_i)
\end{equation}

We generate the following replica to account for y-axis symmetry:
\begin{equation}
\begin{split}
\label{eq:case_3}
(\ref{eq:case_0})\Rightarrow {\chi}_i^{-}=(\tau ,\mu ,\sigma_u^x,\sigma_u^y, \sigma_v^x, \sigma_v^y,\textcolor{blue}{-\theta_u}, \textcolor{blue}{-\theta_v},H_i)
\end{split}
\end{equation}

We account for the non-identifiability of the Gaussian components by swapping components $u$ and $v$ for the cases (\ref{eq:case_3}) and (\ref{eq:case_0}) above:
\begin{equation}
\begin{split}
\label{eq:case_1}
(\ref{eq:case_0})\Rightarrow {\chi}_i^{swap}=( \textcolor{red}{1-\tau} ,\mu , \sigma_{\textcolor{red}{v}}^x,\sigma_{\textcolor{red}{v}}^y,  \sigma_{\textcolor{red}{u}}^x, \sigma_{\textcolor{red}{u}}^y,\theta_{\textcolor{red}{v}},\theta_{\textcolor{red}{u}},H_i)\\
(\ref{eq:case_3})\Rightarrow {\chi}_i^{-swap}=(  \textcolor{red}{1-\tau} ,\mu , \sigma_{\textcolor{red}{v}}^x,\sigma_{\textcolor{red}{v}}^y,  \sigma_{\textcolor{red}{u}}^x, \sigma_{\textcolor{red}{u}}^y,\textcolor{blue}{-\theta}_{\textcolor{red}{v}},\textcolor{blue}{-\theta}_{\textcolor{red}{u}},H_i)
\end{split}
\end{equation}

We also generate replicas to account for the cases of isotropic covariance, where $\sigma_u = \sigma_u^x = \sigma_u^y$ or $\sigma_v = \sigma_v^x = \sigma_v^y$. So, for any data  

\begin{equation}
\label{eq:case_iso_u}
{\chi}_i=(\tau,\mu,\textcolor{orange}{\sigma_u},\textcolor{orange}{\sigma_u},\sigma_v^x,\sigma_v^y,\theta_u,\theta_v,H_i)
\end{equation}
\begin{equation}
\label{eq:case_iso_v}
\textrm{or}\quad{\chi}_i=(\tau,\mu,\sigma_u^x,\sigma_u^y,\textcolor{orange}{\sigma_v},\textcolor{orange}{\sigma_v},\theta_u,\theta_v,H_i)
\end{equation}

We generate replicas

\begin{equation}
\begin{split}
\label{eq:case_iso_u_dup}
(\ref{eq:case_iso_u})\Rightarrow {\chi}_i^{\sigma_u}&=\left\{(\tau,\mu,\textcolor{orange}{\sigma_u},\textcolor{orange}{\sigma_u},\sigma_v^x,\sigma_v^y,\textcolor{orange}{\theta_u},\theta_v,H_i)\right.\\
& \left.|\; \textcolor{orange}{\theta_u}\in \{-\frac{\pi}{2},-\frac{3\pi}{8},-\frac{\pi}{4},-\frac{\pi}{8},0,\frac{\pi}{8},\frac{\pi}{4},\frac{3\pi}{8},\frac{\pi}{2}\}\right\}\\
(\ref{eq:case_iso_v})\Rightarrow {\chi}_i^{\sigma_v}&=\{(\tau,\mu,\sigma_u^x,\sigma_u^y,\textcolor{orange}{\sigma_v},\textcolor{orange}{\sigma_v},\theta_u,\textcolor{orange}{\theta_v},H_i)\\
&| \textcolor{orange}{\theta_v}\in \{-\frac{\pi}{2},-\frac{3\pi}{8},-\frac{\pi}{4},-\frac{\pi}{8},0,\frac{\pi}{8},\frac{\pi}{4},\frac{3\pi}{8},\frac{\pi}{2}\}\}
\end{split}
\end{equation}

The initial data and all its replicas form the extended dataset $\mathcal{X}_{uv}^{all}$: 
\begin{equation}
\begin{split}
\label{eq:case_final_set}
\mathcal{X}_{uv}^{all}=\{{\chi}_i,{\chi}_i^{-},{\chi}_i^{swap},{\chi}_i^{-swap}, {\chi}_i^{\sigma_u},{\chi}_i^{\sigma_v}|{\chi}_i\in \mathcal{X}^{align}_{uv}\}
\end{split}
\end{equation}

Then we filter out any duplicate data from that set to avoid over-sampling of some data and get the  final set  used to train the classifier:

\begin{equation}
\begin{split}
\label{eq:case_final_training_set}
\mathcal{X}^{uni}_{uv}=Unique(\mathcal{X}_{uv}^{all})
\end{split}
\end{equation}

\subsection{Training merging models}

Finally, training on $\mathcal{X}^{uni}_{uv}$, we can obtain the ClustML merger $\mathcal{G}_{ClustML}^*$ optimal at predicting the labels $H_i$ from the input $\phi_i\in\Phi^{uni}_{uv}$, and use it to predict the label $\hat{H}$ of the current input $\phi^{*align}_{uv}$ (\ref{eq:phi_align}): 
\begin{equation}
\hat{H}=\mathcal{G}^*_{ClustML}(\phi^{*align}_{uv})\in{\{0,1\}}
\end{equation}

$\hat{H}$ estimates the unobserved aggregated judgments humans would make for the scatterplot $SP(X\sim \mathcal{M}(\phi^*_{uv}))$.

The training process uses a standard approach in data-driven estimation of parameters of supervised classifiers (see details in Experiment 1). 


\section{Experiments}

ClustML and ClustMe are both GMM-based VQMs. We first demonstrate our claim that the merging decision of ClustML, being trained on perceptual data, is better than the one from ClustMe based on heuristics. ClustMe merging decision \textit{Demp} was the best over six other merging heuristics assessed on the benchmark dataset S1~\cite{ClustMe_eurovis2019}. We use the same dataset S1 to get the optimal merging decision $\mathcal{G}^*_{ClustML}$ for ClustML, and we show that this merging decision is better than $Demp$ on S1, hence, also better than the six other merging heuristics. 

Second, ClustMe VQM has already been proven more accurate than competitors at ranking scatterplots based on cluster patterns on the benchmark dataset S2~\cite{ClustMe_eurovis2019}. We use the same benchmark S2 to show that ClustML VQM improves accuracy over ClustMe VQM and, hence, over previous competitors.


Finally, we propose a usage scenario of ClustML with real genomic data.

\subsection{Experiment 1: training the ClustML merger on perceptual data}
\label{sec:train_ClustML_merger}

To get the ClustML classifier for merging, we first align and augment the data and human judgments from the available dataset, 
then, we present the classification techniques and protocols,  
and finally, select the best among the trained classifiers. 

\subsubsection{Human judgments data}
\label{sec:exp_hum_judg_data}
The initial human judgment data  from study S1 
~\cite{ClustMe_eurovis2019} is summarized in table \ref{tab:Parameters}.
We ignore the $\alpha$ parameter, which gave an additional random rotation to the whole scatterplot for each trial, and we ignore the number $N\in\{100,1000\}$ of points generated in the scatterplot, as none of these parameters appears in the GMM modeling the density of the scatterplot. The dataset $\Phi_{uv}$ is a sample of $1000$ of these scatterplots. We discovered four of them are duplicates, which means four scatterplots were generated twice with the same set of parameters but differing by the number of sampled points ($N=100$ and $N=1000$), or they turned out having the same parameters after data alignment. These four duplicates were removed. We ended up with $996$ unique scatterplots with $34$ human judgments that we summarized by majority vote. 
The final alignment and augmentation processes (equation (\ref{eq:case_final_training_set})) led to $16181$ scatterplots in the set $\mathcal{X}^{uni}_{uv}$ forming the \textit{Benchmark dataset 1}.

\begin{table}
\centering
\caption{Initial data S1 from~\cite{ClustMe_eurovis2019} $\phi_{uv}$ are $1000$ unique parameter sets $\phi_{uv}^{[i]}$ picked randomly among the following values. In this work, we ignore $\alpha$ and $N$. It remains $996$ unique sets of parameters forming $\mathcal{X}^{align}_{uv}$.} 
\resizebox{0.9\columnwidth}{!}{
\begin{tabular}{ | p{1cm} | p{3cm} | p{3.7cm} |}
 \hline
 Param. & Description & Values \\
 \hline
 $\tau$ &  Prior proba. of $u$ &  \{0.1, 0.2, 0.3, 0.4, 0.5\} \\
\hline
$\mu$ &  $u$ to $v$ Euclid. dist. &  \{0, 1, 2, 3, 5, 8, 13, 21\} \\
\hline
$\sigma_{u,v}^{x,y}$ &  Scaling factors  &  \{0.5, 1, 1.5, 2, 2.5, 3\} \\
\hline
$\theta_u$, $\theta_v$ & Rotation angles &	\{0, $\pi/8$, $\pi/4$, $3\pi/8$, $\pi/2$ \} \\
\hline
$\alpha$ &  Rot. angle of $SP(X)$ & \{0, $\pi/2$, $5\pi/4$\} \\
\hline
$N$ &  Num. of points $X$ & \{$100$, $1000$\} \\
\hline
\end{tabular}
}
\label{tab:Parameters}
\end{table}

\subsubsection{Classifiers and training protocol}
\label{sec:exp_classif_CARET}

The $996$ initial scatterplots, although chosen to cover the space of parameters uniformly, were assigned unequally to the two classes by the $34$ S1's subjects. Therefore, $81.5\%$ of the data ended up with a merging decision of $H^{[i]}=1$. This class imbalance requires a specific process for training classifiers to avoid bias favoring the majority class. 
Another issue is the relative scale of the parameters; for instance, the parameter $\mu$ scales up to two orders of magnitude larger than $\tau$. Correlated features must also be dealt with. This requires \emph{pre-processing} steps.

The $16181$ scatterplots $\mathcal{X}^{uni}_{uv}$ were stratified by class, each subset being randomly split into $80\%$ training and $20\%$ testing to finally get $12945$ training and $3236$ test points preserving the (imbalanced) class distribution. Notice that the $3236$ test data points correspond to $709$ of the $996$ unique scatterplots while the $12945$ training data points correspond to $991$ of them. Still, none is duplicated in the parameter space $\mathcal{S}$ after augmentation, forming valid independent training and test sets for learning the automatic classifiers \emph{in that space}.

We used the R-package \texttt{CARET}~\cite{Rpackagecaret} for training $12$ classification techniques, trying $4$ methods to deal with class imbalance, and $4$ pre-processing methods for scaling and remove correlated features, all summarized in Table \ref{tab:CARETparam}. This process resulted in $320$ different classification models.
We used $10$-fold cross-validation on the training set, with $10$ repetitions of the training with random initialization, 

\begin{table}[h!tbp]
\centering
\caption{Methods used from the R-package \texttt{CARET}~\cite{Rpackagecaret,CARET_Kuhn2008}.  
Note \textit{xgbTree} and \textit{gbm} only used \textit{None} and \textit{C} pre-processing.}
\resizebox{0.95\columnwidth}{!}{
\begin{tabular}{|p{0.3cm} |p{2.6cm} |p{4.5cm} |}
\hline
&Method & Description \\ \hline
\multirow{6}{*}{\rotatebox{90}{Pre-processing}} & None & No pre-processing\\
&\textit{Center+Scale} (C) & Zero mean and unit variance\\
&C+\textit{BoxCox} (CB) & Box-Cox transformation\\
&C+\textit{PCA} (CP) & Principal Component Analysis \\
&C+B+P (CBP)& \\
&CBP+\textit{spatialSign} (CBPS) & Dividing by norm (unit sphere)\\ 
\hline
\multirow{5}{*}{\rotatebox{90}{Balancing}} & None & No balancing \\
&upSample & Rand. replica of minor. class \\
&downSample & Rand. sampling of major. class \\
&ROSE& Rand. over-sampling~\cite{ROSE_Menardi2014}  \\
&Smote & Synth. minor. class NN~\cite{SMOTE_Chawla2002}\\\hline
\multirow{12}{*}{\rotatebox{90}{Classification technique}} 
&nb    & Naive Bayes\\
&knn   & k-Nearest Neighbors\\
&rf     & Random Forest\\
&treebag & Bagged Classif. Adap. Reg. Tree \\
&blackBoost & Boosted Reg. Tree \\
&gbm  & Gene. Boosted Reg. Model\\
&xgbTree & Extreme Gradient Boosting Tree\\
&earth & Multivar. Adap. Reg. Spline\\
&svmRadial    & Radial Kernel Sup. Vec. Mach. \\
&mlpWeightDecay & Multi-Layer Perceptron \\
&glm & Generalized Linear Model \\
&glmnet &  GLM penal. max. lik.\\
 \hline
\end{tabular}
}
\label{tab:CARETparam}
\end{table}

To evaluate and select the best classifier on the test data, we computed the \textit{Matthews Correlation Coefficient} (MCC), which is regarded as immune to large class imbalance~\cite{bekkar2013evaluation,MCC_Boughorbel2017}.

\begin{figure}[h!tbp]
\begin{center}
\resizebox{0.9\linewidth}{!}{
\begin{tabular}{ccc}
\includegraphics[width=0.3\linewidth, height=5cm]
{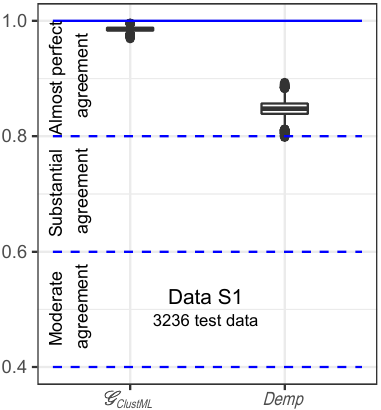}
&\includegraphics[width=0.3\linewidth, height=5cm]
{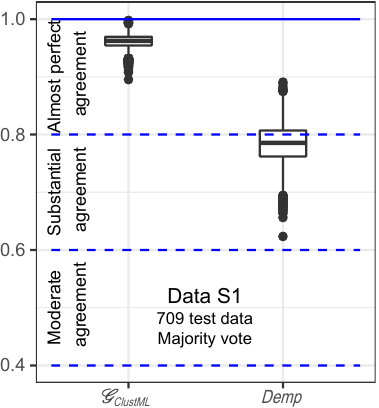}
&\includegraphics[width=0.3\linewidth, height=5cm]
{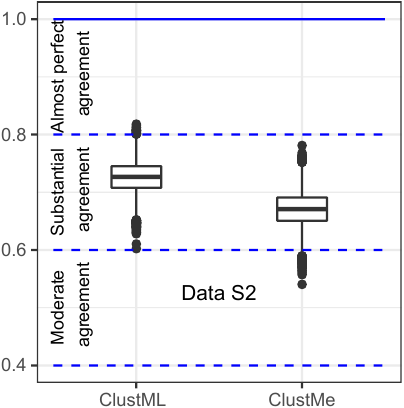}\\
(a) Experiment 1& (b) Experiment 1 & (c) Experiment 2
\end{tabular}
}
\caption{
ClustML merger ($\mathcal{G}_{ClustML}$) is noticeably better than ClustMe merger ($Demp$) based on Vanbelle Kappa score on the $3236$ augmented data of the test set in \textit{Experiment 1} (a) and on the $709$ test scatterplots with class computed by majority vote of $\mathcal{G}_{ClustML}$ predictions on augmented data (b). ClustML VQM is noticeably better than ClustMe VQM at ranking the $435$ pairs of scatterplots in \textit{Experiment 2} (c). However, as expected, scores are lower than in Experiment 1 as these scatterplots display more complex patterns and involve the full VQM pipeline.}
\label{fig:clustme_ClustML_vanbelle_bootstrap_exp1and2}
\end{center}
\end{figure}

\subsubsection{ClustML merger is better than Demp.}
\label{sec:exp_result_training}

Table \ref{tab:classifiersResults} lists the best setting for each classification technique. 
The overall best combination to realize the \emph{ClustML merging function}  $\mathcal{G}_{ClustML}$ is a bagged Classification and Regression Tree (CART) model (\texttt{treebag}) with up-sampling of the minority class (\texttt{upSample}) and running all pre-processing methods (\texttt{Center + Scale + BoxCox + PCA + spatialSign}). In study S1~\cite{ClustMe_eurovis2019}, $Demp$ is the best among seven merging heuristics and is used to form $ClustMe$.
Following that study, we use the Vanbelle's Kappa $\kappa_{v}$ agreement index~\cite{VanbelleKappa_2009} to compare both merging techniques with the $34$ human judgments.

\begin{table}[h!tbp]
\centering
\caption{The best of each classification technique based on Matthew's correlation coefficient (MCC) is given together with its specific class balancing and pre-processing compounds (See Table \ref{tab:CARETparam}). Treebag with upsampling and all pre-processing options is the most accurate on the $3236$ test data of \textit{Experiment 1}.}  

\resizebox{0.8\columnwidth}{!}{
\begin{tabular}{|c|c|c|c|}
\hline
Classification tech.  & Balancing & Pre-Processing & MCC\\
\hline
treebag & upSample & CBPS & \textbf{0.970}\\
\hline
rf & None & None &	0.959\\
\hline
gbm & upSample & None	& 0.953\\
\hline
mlp & None & CBPS	& 0.893\\
\hline
knn & None & CB &	0.888\\
\hline
earth & None & CBPS &0.876\\
\hline
blackBoost  & Smote & C &	0.875\\
\hline
xgbTree & None & None &	0.868\\
\hline
svmRadial & None & CBPS &	0.866\\
\hline
glmnet & None &  CBPS &	0.832\\
\hline
glm & None & None  &	0.831\\
\hline
nb  & None & CB	& 0.828\\
\hline
\end{tabular}}
\label{tab:classifiersResults}
\end{table}

Vanbelle's kappa $\kappa_v$ considers both the agreement between the group of human raters and the VQM and the within-group inter-rater agreements. The $\kappa_v$ values are interpreted using a standard scale~\cite{landis1977measurement}: $<0$ \emph{poor}, $]0,0.2]$ \emph{slight}, $]0.2,0.4$ \emph{fair}, $]0.4,0.6]$ \emph{moderate}, $]0.6,0.8]$ \emph{substantial}, and $]0.8,1]$ \emph{almost perfect} agreements. We run $10000$ evaluations on the bootstrap samples~\cite{bootstrap_efron93} of the test data to estimate the average score the two mergers would have obtained varying the distributions of scatterplot parameters and to better quantify their difference.

There are two ways to compare $\mathcal{G}_{ClustML}$ and $Demp$ mergers. In \textbf{case 1}, we compute $\kappa_{v}$ on the $3236$ augmented test data, which are unique for $\mathcal{G}_{ClustML}$ but duplicates of some of the $709$ $Demp$ merging decisions, biasing the comparison towards the duplicate cases (Figure \ref{fig:clustme_ClustML_vanbelle_bootstrap_exp1and2} left). In \textbf{case 2}, we compute $\kappa_{v}$ on the $709$ scatterplots from the test set, which is fair for $Demp$, but forces us to summarize the $\mathcal{G}_{ClustML}$ predictions by a majority vote over the duplicated data. (Figure \ref{fig:clustme_ClustML_vanbelle_bootstrap_exp1and2} center). 

In case 1 favoring $\mathcal{G}_{ClustML}$, it gets $\kappa_{v}=0.986$, $16\%$ greater than $Demp$'s $\kappa_{v}=0.848$, both being in \emph{Almost perfect} agreement with human judgments.
In case 2, favoring $Demp$, $\mathcal{G}_{ClustML}$ gets $\kappa_{v}=0.962$ (\emph{Almost perfect} agreement), a $22\%$ improvement over $Demp$'s $\kappa_{v}=0.786$ (\emph{Substantial} agreement; consistent with the state-of-the-art score $\kappa_{v}=0.788$ computed over the full $1000$ dataset~\cite{ClustMe_eurovis2019}). 
 The ClustML merger ($\mathcal{G}_{ClustML}$) is better than the ClustMe merger ($Demp$) by a large margin, with more than $15\%$ accuracy improvement in both cases.

\subsection{Experiment 2: ClustML is better at ranking scatterplots}
\label{sec:exp_result_ranking}

In this experiment, we compare ClustML and ClustMe. Both are GMM-based VQMs, as illustrated in Figure \ref{fig:ClustML_pipeline}a.
We use them to rank pairs of scatterplot projections of real and synthetic multidimensional data from the dataset S2~\cite{ClustMe_eurovis2019}. None of these scatterplots has been used in the training process of \mbox{ClustML}, nor in determining parameters of ClustMe.
S2 is made of all $435$ possible pairs of $30$ monochrome scatterplots selected among the dataset composed of $257$ scatterplots from an earlier study~\cite{taxonomySepme_SedlmairTMT12}. $ 31$ subjects have judged each pair to rank the scatterplots by the perceived group structure complexity of the displayed point patterns on a 3-category scale: ``$<$'', ``$=$'', ``$>$''. 

We use the \texttt{mclust} R-package with $BIC$ model selection to train the GMM. We run ClustML and ClustMe merging functions on each pair of components identified by the GMM and finally get the respective VQM for each of the $30$ scatterplots. Finally, following~\cite{ClustMe_eurovis2019}, we use this VQM score to rank the scatterplots, and we compare the ranking with that of human judgments on all $435$ pairs using the Vanbelle's kappa index.

ClustML gets $\kappa_v=0.727$, improving over ClustMe's $\kappa_v=0.671$ top score to date.
Figure \ref{fig:clustme_ClustML_vanbelle_bootstrap_exp1and2} shows the $10000$ bootstrap samples distribution of the $435$ pairs of scatterplots for the two VQMs. ClustML is still noticeably better than ClustMe on this data. However, the score difference is lower than in the previous experiment with only $8\%$ improvement, and both scores are within the \textit{Substantial agreement} range.

\subsection{Qualitative comparison of ClustML and ClustMe}

\begin{figure}[h!tbp]
\centering
\begin{tabular}{c}
\includegraphics[width=0.6\columnwidth]{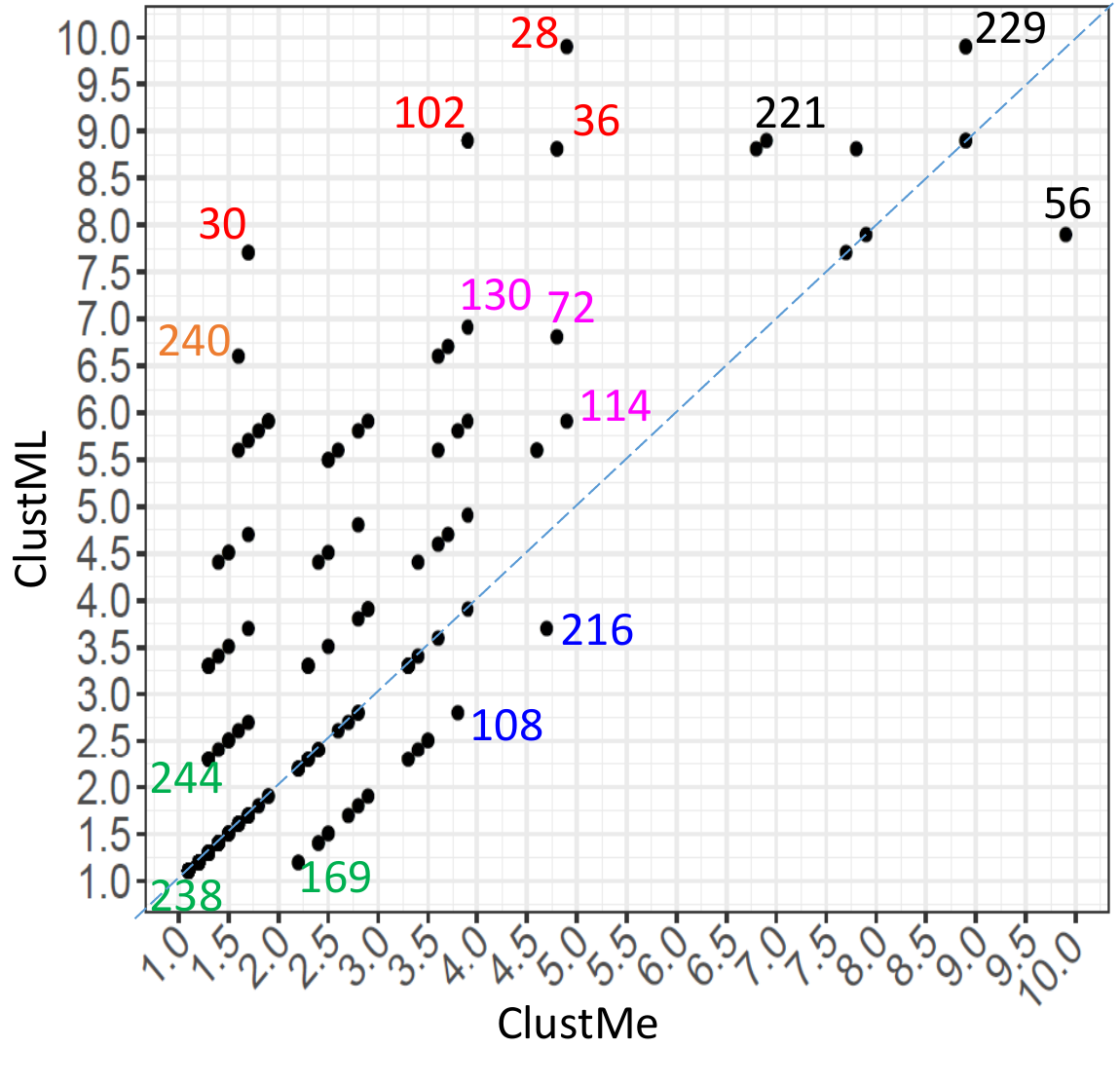}\\
\includegraphics[width=0.6\columnwidth]{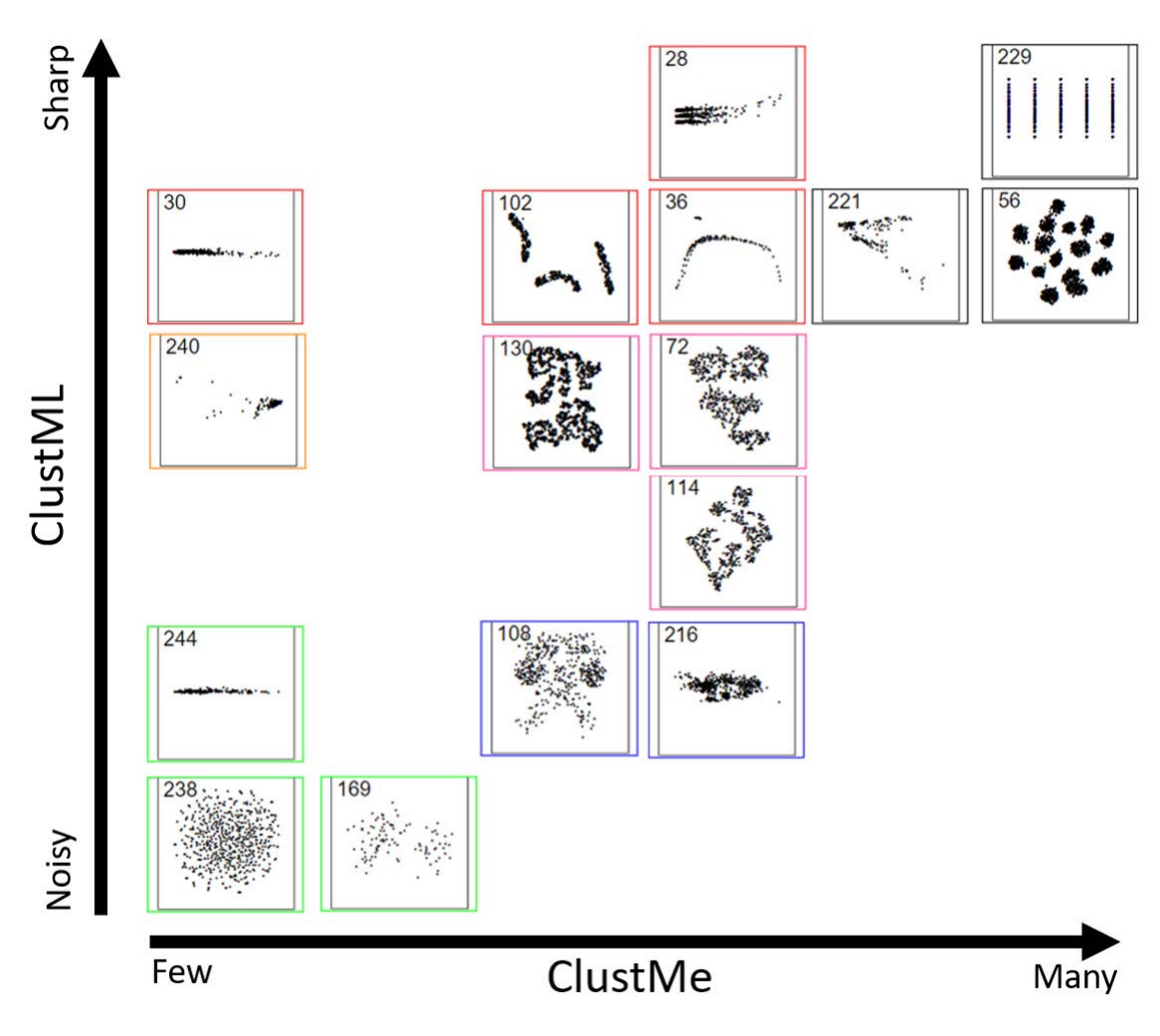}
\end{tabular}
\caption{Top: Comparison of ClustML and ClustMe scores of 257 scatterplots from~\cite{taxonomySepme_SedlmairTMT12}. Bottom: $16$ selected scatterplots (SPs) from the top view with corresponding colors, numbers, and approximate locations.
Both ClustMe and ClustML give equally low scores to SPs $\textcolor{green}{238, 244, 169}$ with no strong cluster patterns and similarly high scores to SPs $\textcolor{black}{221, 229, 56}$ with sharp and numerous cluster patterns.
ClustML gives high scores to SPs \textcolor{red}{28, 36, 102}, medium scores to SPs \textcolor{magenta}{72, 114, 130}, and low scores to SPs \textcolor{blue}{108, 216}, while ClustMe gives them all a medium score. ClustML seems better than ClustMe at distinguishing \textcolor{red}{sharp} cluster patterns from \textcolor{magenta}{slightly noisy} and \textcolor{blue}{very noisy} ones. 
On the other hand, ClustMe seems more sensitive to the cluster numerosity, distinguishing low numerosity clusters in SPs \textcolor{red}{30} and \textcolor{orange}{240} from medium numerosity in SPs \textcolor{magenta}{72, 114, 130}, and high numerosity in SPs \textcolor{black}{229, 56} while ClustML gives to all of them a medium-high score.}
\label{fig:benchmark3_ranking257}
\end{figure}

ClustMe and ClustML are used to rank the $257$ scatterplots from~\cite{taxonomySepme_SedlmairTMT12}. Their scores are compared in Figure \ref{fig:benchmark3_ranking257}. 
Scatterplots (SPs) at the bottom show details of the dots in the top view. Numbers indicate identifiers of the SPs in the dataset. The caption of the figure gives detailed observations. 
ClustML seems more sensitive to cluster sharpness, while ClustMe seems more sensitive to cluster numerosity.

\subsection{Interpretation of Vanbelle kappa with a worst-case analysis}
\label{sec:worstcase_analysis}

\begin{figure}[h!tbp]
\begin{center}
\includegraphics[width=0.7\linewidth]{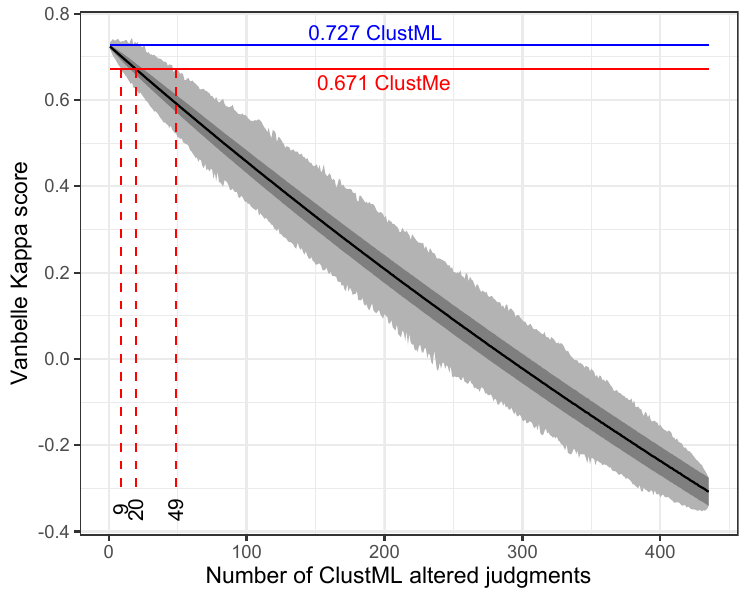}
\caption{Distribution of Vanbelle kappa when altering $10000$ time $k$ values randomly chosen among the ClustML decisions ($k\in \{1,\dots,435\}$) over the $435$ pairs of $30$ scatterplots in \textit{Experiment 2}.  
The dark grey area shows one standard deviation above and below the average value (black line). Light grey extends between the minimum and maximum values of the $10000$ samples. This serves to evaluate how much ClustMe would worsen ClustML ordering.}
\label{fig:clustme_ClustML_altered_vanbelle_bootstrap_exp2}
\end{center}
\end{figure}

To better understand the meaning of these ranking scores, we compute the Vanbelle index when altering $10000$ times, $k$ ClustML decisions for each $k\in\{1,\dots,435\}$ randomly. By alteration, we mean changing any of $<$, $=$, or $>$ order relations to a different order relation from the same set. The resulting distribution of the Vanbelle Kappa for each value of $k$ is displayed in Figure \ref{fig:clustme_ClustML_altered_vanbelle_bootstrap_exp2}. The ClustML score decreases in proportion to the number of alterations. It requires between $r_{min}=9$ ($2\%$) and $r_{max}=49$ ($11\%$) alterations, with $r=20$ ($4.6\%$) on average, to get down to the ClustMe score. 

Altering a decision occurs whenever the order of two of the scatterplots is changed. For instance, on average, the difference between ClustML and ClustMe is equivalent to moving a single scatterplot down or up by $20$ positions in the total ordering or changing the rank of more scatterplots by a total of $20$ rank alterations. 

Let us consider a realistic usage scenario where the user has a \textit{time budget} so they can afford to explore only the top-K scatterplots in depth in search of new insights. 
Moving $n$ elements out of the top-K set ($K\geq n$) requires at least $r=n^2$ rank permutations if we pick up the bottom $n$ of that set. For instance, if $\bm{abcdef}|ghijkl...z$ is an ordered set of $26$ scatterplots and the user as a time budget to explore only the top $K=6$ ($\bm{a}$ to $\bm{f}$ delimited by $|$), then moving $n=2$ scatterplots out of the top-6, say $\bm{e}$ and $\bm{f}$ to get $\bm{abcd}gh|\bm{ef}ij...z$, requires altering at least $r=4$ pairwise rankings ($g\leftrightarrow e$,\;$g\leftrightarrow f$,\;$h\leftrightarrow e$,\;$h\leftrightarrow f$). To push any set of $n$ items out of the top-K, any other group of permutations requires \textit{at least} $n^2$ rank permutations. 
Hence, in the worst case, given a ClustML ordering of the $30$ scatterplots, ClustMe, in comparison, may down-rank between $n_{min}=\sqrt{r_{min}}=3$ and $n_{max}=7$ scatterplots off the top-$K$ most potentially insightful ones. It is also possible that most or all the alterations created by ClustMe occur outside of the top-K, so they would not impact the time-budgeted insight gathering, ClustMe and ClustML having identical top-K sets.

We can extrapolate this observation to any dataset size. 
For a dataset with $N$ scatterplots, a simple calculus shows that if $p$ is the \textit{percentage of ranking alterations} over the $N(N-1)/2$ pairs, with $p\leq 50$, then the \textit{number of ranking alterations} is $r=N(N-1)p/200$. Moreover, in the worst case, the \textit{percentage of down-graded scatterplots} $q$ is $100\sqrt{r}/N$. Hence, $q= 10\sqrt{(N-1)p/2N}\approx \sqrt{50p}$ for large $N$. 

ClustMe alters in the worst case about $11\%$ of all pairs of $N$ scatterplots ($p=100\times r_{max}/N=100\times 49/435\approx11$). Therefore, it could downgrade up to $23.5\%$ of the scatterplots ordered by ClustML in a worst-case scenario.

\section{Usage scenario with genomic data}

To illustrate ClustML's potential utility to real-world analyses, we provide a usage scenario with genomic data.
In many domains of micro-biology, analysts rely on data visualization to spot interesting patterns that deserve further detailed analysis. Automatic clustering of single-cell data is known to be challenging~\cite{Kiselev2019}. 
As such, biologists often resort to dimensionality reduction and visualizing scatterplots to decide about clusters of cells and their features~\cite{Feng_2020}. Alternatively, scatterplot matrices (SPLOMs) are used, for instance, to visually identify interesting groups of cells in scatterplots determined by pairs of eigengenes (axes), each eigengene coding a group of coexpressed genes~\cite{Han2017}.
In genome-wide association studies, analysts project the genetic data into principal components space for visual inspection~\cite{KinVisPvis_Aupetit2016,KinVisSoftware_Ullah2019}. In all these situations, the numerous projection methods and their parameters lead to possibly hundreds of scatterplots representing different facets of the same multidimensional data, similar to the type of data used in study S2.  

\begin{figure}[h!tbp]
\begin{center}
\begin{tabular}{cc}
 \multicolumn{2}{c}{\includegraphics[width=0.8\linewidth]{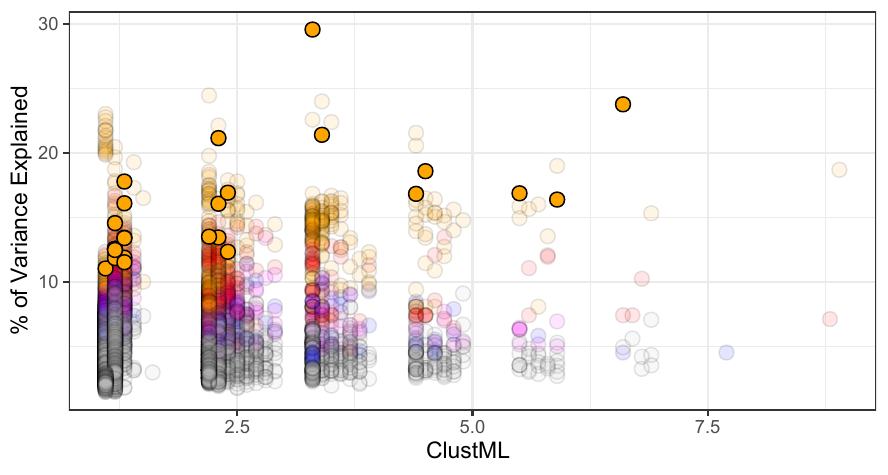}}\\
\includegraphics[width=0.45\columnwidth]{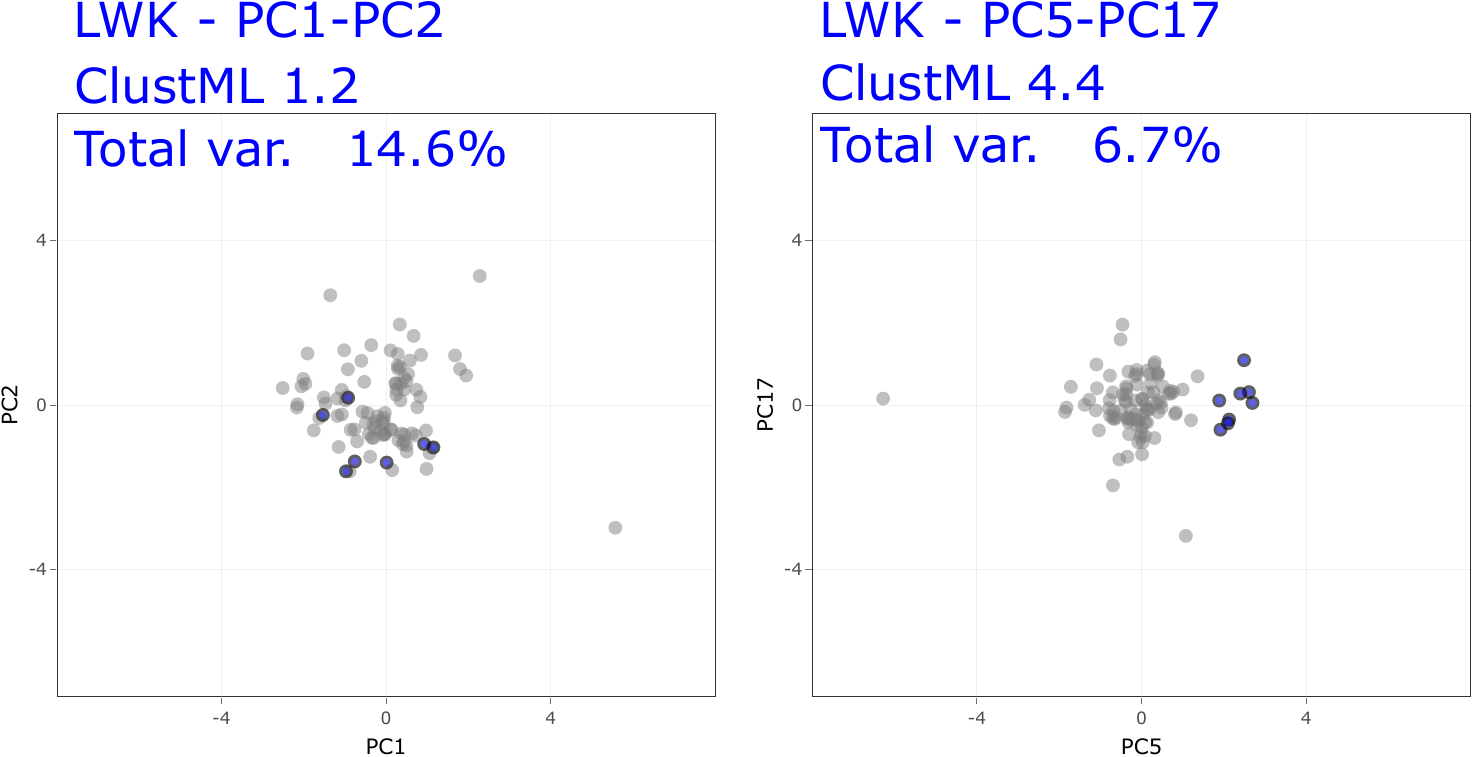}&
\includegraphics[width=0.45\columnwidth]{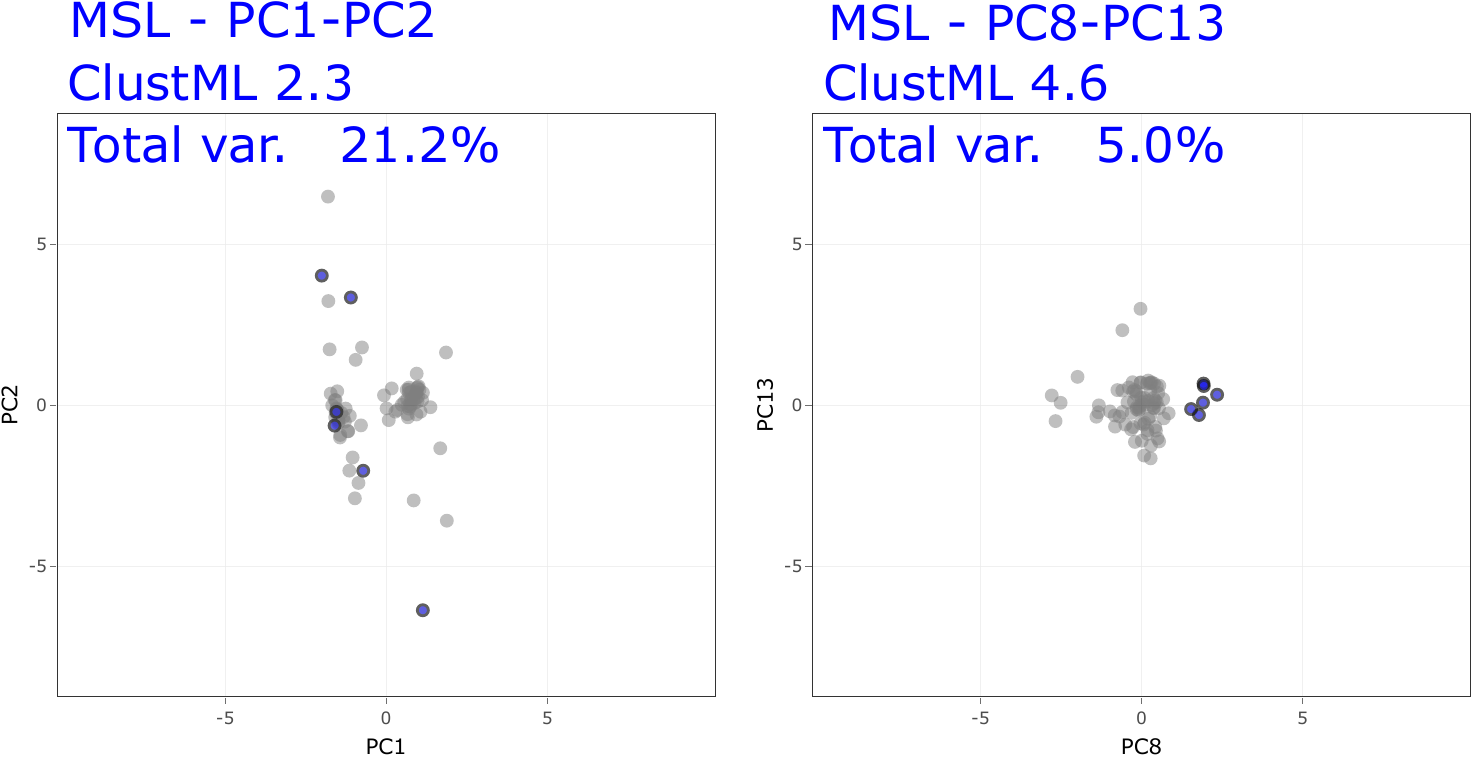} \\
\includegraphics[width=0.45\columnwidth]{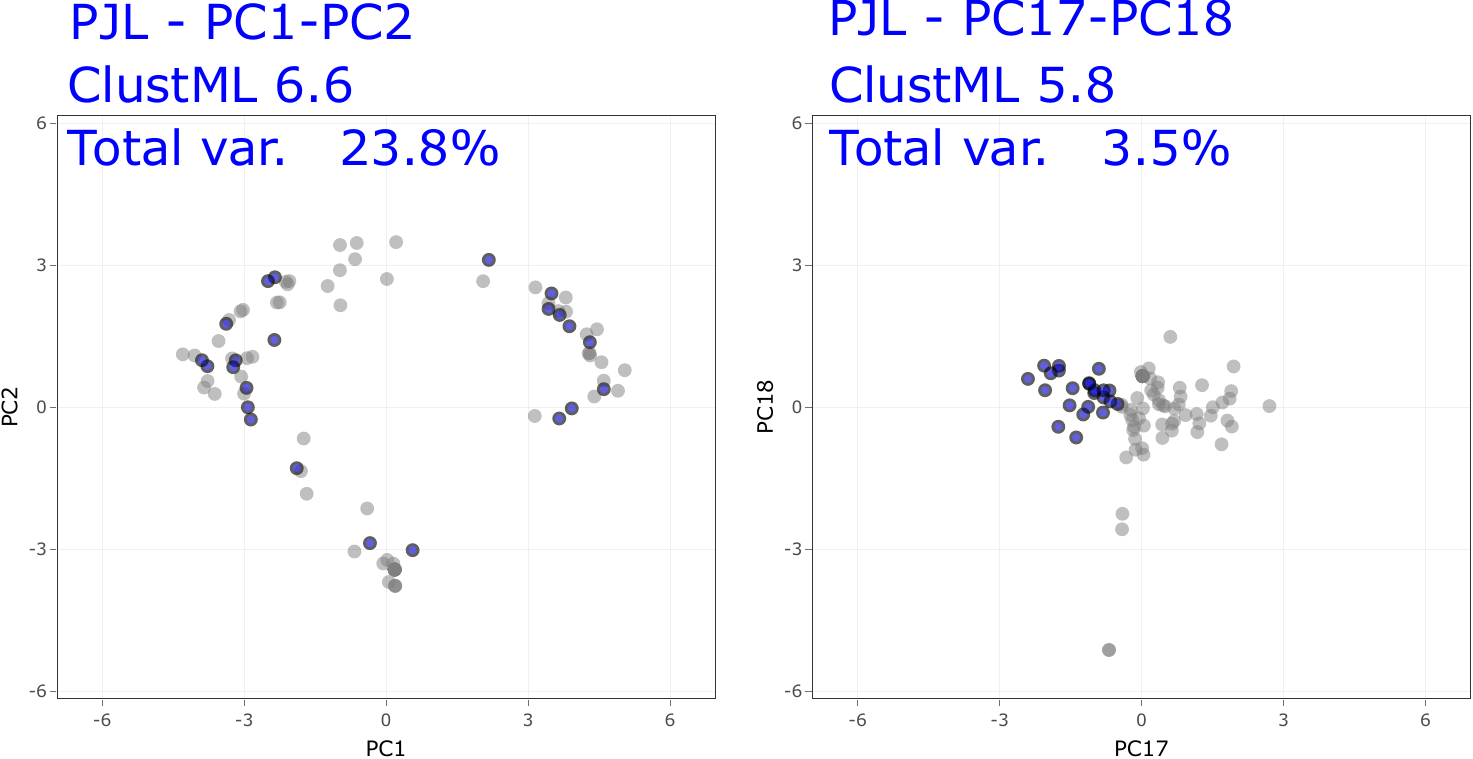}&
\includegraphics[width=0.45\columnwidth]{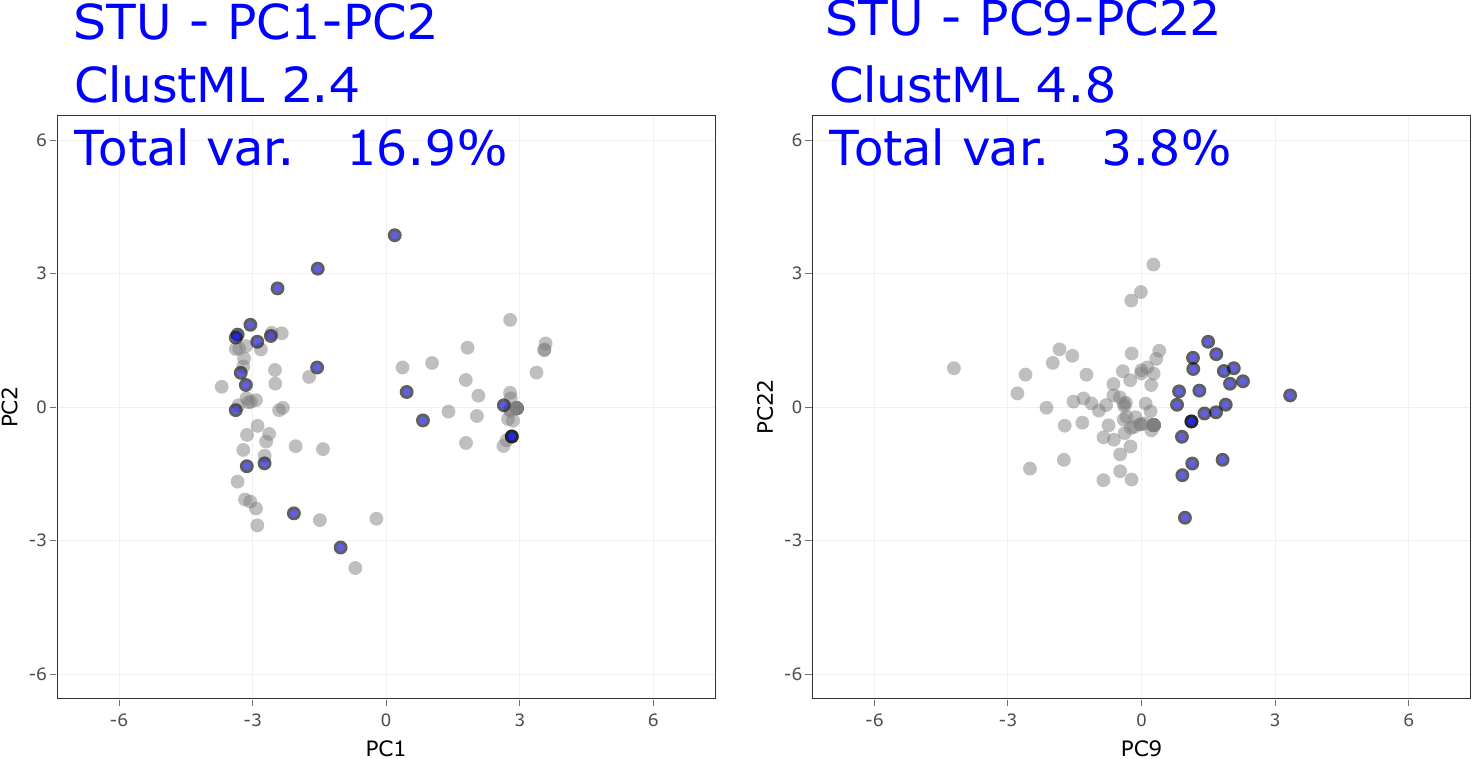}
\end{tabular}
\caption{Top: Distribution of 11310 scatterplots from all pairs of top 30 principal components (PCs) of 1000 Genome Project kinship data in the space of ClustML score and percentage of variance explained. Dots are color-coded by the axis with the most variance in the scatterplots, showing the ones directed mainly by the \textcolor{orange}{first}, \textcolor{red}{second}, \textcolor{magenta}{third}, or \textcolor{blue}{fourth} PC (black otherwise). Solid orange dots with thick edges are SPs directed by the first and second PCs. Analysts typically limit their exploration to SPs, explaining most of the variance at the top of the summary scatterplot. Those scatterplots mostly involve top PCs only. 
Bottom: we show top-level SPs directed by PC1-PC2 of LWK, MSL, PJL, and STU populations (left side of each pair), and lower-level SPs directed by PC5-PC17 for LWK, PC8-PC13 for MSL, PC17-PC18 for PJL, and PC9-PC2 for STU (right side of each pair). The lower-level SPs have about the same or even a higher ClustML score than the PC1-PC2 SPs of the same population. ClustML allows the analyst to detect a cluster pattern in each population (manually lassoed blue dots, right side), which would have been missed exploring only the PC1-PC2 SP of that same population (left side) because the same data points do not form a cluster pattern therein. Notice pairs of SPs are displayed at the same scale, showing that cluster patterns on their right side are of similar importance to those on their left side in terms of within and between variance.}

\label{fig:SPLOM_scRNAseq}
\end{center}
\end{figure}

In this usage scenario, we consider the data from the $1000$ Genome Project phase 3 dataset~\cite{Auton2015} composed of genetic data of $26$ populations of about $100$ individuals each. We measure kinship between individuals of each population separately, computing identity-by-descent~\cite{KinVisSoftware_Ullah2019}. We project these data using Multidimensional Scaling into $30$ dimensions and compute ClustML on each possible pair of principal components for each of the $26$ populations separately. The top view in Figure \ref{fig:SPLOM_scRNAseq} shows the $11310$ SPs in the space of the ClustML score and the proportion of variance explained. The bottom view shows several SPs found exploring the highest ClustML scores in search of complex patterns that could relate to subgroups of individuals in each population. 

Analysts typically rely on exploring SPs spanning pairs of the top-most principal components only (orange dots with a thick edge), possibly missing essential patterns as pointed out in a recent work~\cite{Elhaik2022}. Thanks to ClustML, we can discover SPs spanning lower order components (\textit{e.g.} down to the 17th PC for the PJL population) containing cluster patterns of potential interest to the analyst which cannot be detected in the SPs directed by the top two principal components (See Figure \ref{fig:SPLOM_scRNAseq} bottom). ClustML can guide the analyst, avoiding a very costly exhaustive exploration of the $11310$ SPs.

\section{Discussion and future work}

We proposed a new data-driven, GMM-based VQM for cluster patterns. ClustML's main novelty is to use a merging component fully trained on human judgment data.

\textbf{Options for improving ClustML} The ClustML merging component uses a majority vote to transform the collective judgments of $34$ participants into a binary value, losing the richness of the human judgments more akin to a probability value. The GMM also restricts the type of cluster patterns that can be quantified to a mixture of Gaussians while other types of distributions and mixtures could be explored.  
At last, GMM-based VQMs act at the geometric encoding stage of the visualization pipeline, ignoring the aesthetic aspect of the scatterplot like color, opacity, size, and shape of the marks; other parameters which can also impact the perception of cluster patterns~\cite{Quadri2021,Quadri2022}. All these aspects leave room for further study and improvements.

\textbf{Towards hybrid computational-perceptual models of cluster patterns} It is typically challenging to learn a model for usually unsupervised tasks such as cluster pattern quantification: there is a lack of available representative and human-annotated scatterplot data to train supervised models due to an extreme variation of the cluster patterns~\cite{taxonomySepme_SedlmairTMT12, VisualClusterAnalysis_PandeyKFBB16}, and a lack of a relevant representation space common to all these data. A related approach uses a deep network model~\cite{ScatterNet_Ma2018} trained on scatterplot images to model the human-perceived similarity between patterns in monochrome scatterplots; working with the image pixels as common representation space is an ecologically valid option but still requires collecting enough human-annotated data to cover the vast amount of possible patterns in visualization images. Other heuristic-based techniques use a binning process to reduce the dimension of the image space where to look for visual patterns~\cite{Quadri2021,Quadri2022}. In contrast, in this work, we transformed a typically unsupervised cluster pattern quantification problem into a supervised one, observing that the GMM (Stage 1) acts as a representation model, embedding the underlying points ${X}$ of a scatterplot $SP(X)$ into the GMM's parameter space. By considering only pairs of GMM components, this representation space additionally got a fixed and reduced dimensionality, not only enabling the use of standard supervised classifiers but also drastically limiting the variety of cluster patterns to be learned (two-Gaussian-based distributions only), so the amount of data to be collected. Finally, it happened that the scatterplot stimuli of the S1 dataset were also generated by sampling such a space; hence, they could be used to train such classifiers. This option was not technically straightforward, as demonstrated by the data pre-processing, cleaning, and augmentation steps required to train the ClustML's merging function. Overall, our work opens the door to introducing human-perceptual judgment data in originally unsupervised models, developing new hybrid computational-perceptual models for 
 pattern recognition in visualization and pursuing pioneering work in that area~\cite{PerceptionBasedClustering_AupetitSABB19, SupDRSepMe_Wang2018, OptimColorSepMe_Wang2019}.

The development of such hybrid models raises the question of how to collect a sufficient amount and quality of perceptual data in the first place. Pattern recognition models have long been studied and trained on \emph{natural} images annotated by experts or crowdsourcing~\cite{Irshad2015}. But only a few studies use perceptual-data-driven approaches for pattern recognition in \emph{visualization} images ~\cite{Sepme1_SedlmairA15,Sepme2_AupetitS16,ScatterNet_Ma2018}. We advocate for driving new research in that area to develop data-driven perceptual-based VQM for clusters and other visual patterns in scatterplots, parallel coordinate plots, and other visualization idioms  ~\cite{Dasgupta2010Pargnostics,qualMetric_Bertini11}. 


\textbf{Beyond user study evaluations} 
As stated in ~\cite{Munzner2009nestedModelEval,Meyer2012TheFN}, new algorithms like ClustML should typically be evaluated for accuracy and computing resources. But VQMs algorithms are designed to support humans by replacing them in repetitive perceptual tasks ~\cite{VQM_BertiniS06}. Thus, accuracy is measured by comparing VQM scores to human judgments on the same visual stimuli. Hence, the design of new VQMs naturally relies on collecting perceptual judgment data from quantitative user studies. 
However, when the same judgment data can be re-used for comparing different VQM algorithms because they target the same perceptual task, it is unnecessary to run a new user study for each new VQM variant.  
Re-using study data was first achieved successfully for the design and evaluation of data-driven VQMs for class separation in scatterplots~\cite{Sepme1_SedlmairA15,Sepme2_AupetitS16}, with data from an earlier project~\cite{Sedlmair2013_ScatterplotAndDR}.
The present paper is a renewed demonstration of that approach, comparing ClustML with ClustMe on S1 and S2 previously collected study data, relieving us of the need to run another user study to evaluate ClustML. 

The use of benchmark data for algorithmic assessment is standard in computer science~\cite{Munzner2009nestedModelEval} and benefits the replicability, fairness, and objectivity of the comparison while scaling up the design process of new techniques~\cite{Sim2003benchmarkStudy}. Benchmark data also enables the data-driven design of new models using machine-learning techniques. Benchmarking in visualization is not new for comparing algorithmic approaches~\cite{Espadoto2021quantSurveyDR}, but it is pretty novel when considering human judgment data. Once a benchmark of judgment data is set, it avoids investing unnecessary expert resources to design user studies and collect similar data, and it prevents the additional risk of failure in doing so. By being able to re-use previously collected judgment data S1 and S2, our work demonstrates that it is possible to generate such benchmark data once and use them multiple times for the evaluation and the design of new VQMs. Hence, we advocate for including in the design process of quantitative user studies a reflection on the possibility to re-use the collected data \textit{beyond evaluation}, to \textit{enable generating and training new models}. How to develop such benchmark judgment data to facilitate their re-use in visualization design is a challenging research topic worthy of investigation.

\section*{Acknowledgments}
Michael Sedlmair is funded by the Deutsche Forschungsgemeinschaft (DFG, German Research Foundation) -- Project-ID 251654672 -– TRR 161. The Authors declare that there is no conflict of interest.

\end{document}